%% file: jasa_unblinded_version-revision.tex
\documentclass[12pt]{article}
\usepackage[english]{babel}
\usepackage{epsfig}
\usepackage[active]{srcltx}
\usepackage{graphicx, amsmath}
\usepackage{lscape,amsmath,amsxtra}
\usepackage{latexsym}
\usepackage{bm}
\usepackage{bbm}
\usepackage{caption}
\usepackage{JASA_manu}

\usepackage{color} 

\usepackage{natbib}

\usepackage[latin1]{inputenc}
\usepackage{verbatim}
\usepackage{amsfonts}
\usepackage{rotating}
\usepackage{tabularx}

\newcommand{\E}{\mathrm{E}}

\newcommand{\Var}{\mathrm{Var}}

\title{Hierarchical space-time modeling of  asymptotically independent exceedances  with  an application to precipitation data}
\include{authors}
\date{\today}

\begin{document}

\maketitle
\input{abstract}

\noindent%
\vfill
\newpage
\input{introduction}

\input{hierarchical-model}
\input{space-time}

\input{dependence}
\input{composite-likelihood}

\input{simulation}

\input{real-data-example}

\input{conclusions}

\input{acknowledgment}

 \input{appendix}
 \newpage
\bibliography{gamma}
\bibliographystyle{ECA_jasa}

\newpage
\begin{table}[b!]
	\begin{center}
		\begin{tabular}{ccccccc}
			& \multicolumn{6}{c}{Parameters}	\\
			Scenario &  $\gamma_1$& $\gamma_2$ & $\phi$ &  $\delta$ & $\omega_1$ & $\omega_2$\\
			A & 0.2 & 0.2 &-& 10 & 0.00 & 0.00 \\
			B & 0.2 & 0.3 &$\pi/4$& 5 & 0.05 & 0.10 \\
		\end{tabular}
	\end{center}
	\caption{Design of the two simulation scenarios.}\label{tab:sim}
\end{table}

\clearpage
\begin{table}[b!]
	\begin{center}
		\begin{tabular}{lrrrrrrr}
			Model & \multicolumn{5}{c}{Parameters} & \\
 & {$\gamma_1$} & {$\gamma_2$}  & {$\phi$} &  {$\delta$} & $\omega_1$ & $\omega_2$&CLIC$^*$
			\\
G1&		165.062 &  318.823 &    0.085&    20.184&     0.723 &    0.446&	404480.8\\
& \textit{23.459} &   \textit{19.811}  &   \textit{0.026}  &   \textit{0.948} &    \textit{0.195} &    \textit{0.009} &\\
G2 &175.817&  294.323&    0.041&   20.036& 0 & 0 &404488.1\\
& \textit{11.879} &  \textit{25.291}&    \textit{0.064}&    \textit{1.039}& - &-&\\ 
		\hline	
			C1 & $\tau_1$ & $\tau_2$ & $\psi_S$ & $\psi_T$ & &		&CLIC$^*$\\
& 0.057 &  2.568 &  137.692 & 10.128& & & 404626.2\\
&\textit{0.060}&  \textit{0.309}&  \textit{7.615}&  \textit{0.523}& \\
					\hline	
&			$\tau_1$ & $\tau_2$  & $\psi_S$ & & $\nu_1$ &$\nu_2$		&CLIC$^*$\\
			C2 & 1.034    & 2.025       & 108.755 &&6.672 &   16.358& 404750.3 \\
			& \textit{0.040}  &   \textit{0.318}  & \textit{7.299} & & \textit{0.908}   &  \textit{1.502} &\\    
			C3 & 0.481 &    5.125&        174.980& & 6.614   & 10.406 & 405020.4\\ 
& \textit{0.005} &    \textit{0.262} &    \textit{6.955}	&&		  \textit{0.095}&     \textit{0.226} &\\  
		\end{tabular}
		
	\end{center}
	\caption{Estimates, standard errors (in italic) and CLIC$^*$ values  of fitted models.  Parameter units are kilometers for $\phi_S$, $\gamma_1$ and $\gamma_2$, radians for $\phi$ and $\tau_1$, hours for $\delta$ and $\phi_T$ and kilometers per hour for $\omega_1$, $\omega_2$, $\nu_1$ and $\nu_2$.}\label{tab:results}
\end{table}

\clearpage
\begin{table}
	\begin{center}
		\begin{tabular}{lrrrrrr}
			&		\multicolumn{2}{c}{$\mathrm{RMSE}(0)$} & 		\multicolumn{2}{c}{$\mathrm{RMSE}(1)$} &
			\multicolumn{2}{c}{$\mathrm{RMSE}(2)$}
			\\
			
			&		$q_{0.99}$ &  $q_{0.995}$&
			$q_{0.99}$ & $q_{0.995}$&$q_{0.99}$ & $q_{0.995}$\\
			\hline
			G1 & 2.614 & 2.096 & 1.901   & 1.643  & 1.475& 1.496\\

			G2 &2.605 &  2.072  & 1.907 & 1.626 & 1.477& 1.480\\

			C1 & 2.240 & 2.455 	&2.053 	& 2.428& 1.779 & 1.928\\
		\end{tabular}
	\end{center}
	\caption{Total root mean squared errors for the estimates of $\chi^*_{s_i;h}(q)$.}\label{eq:multi-chi}
\end{table}

\clearpage
\begin{figure}
	\begin{center}
		\begin{tabular}{cc}
			\includegraphics[width=0.4\textwidth]{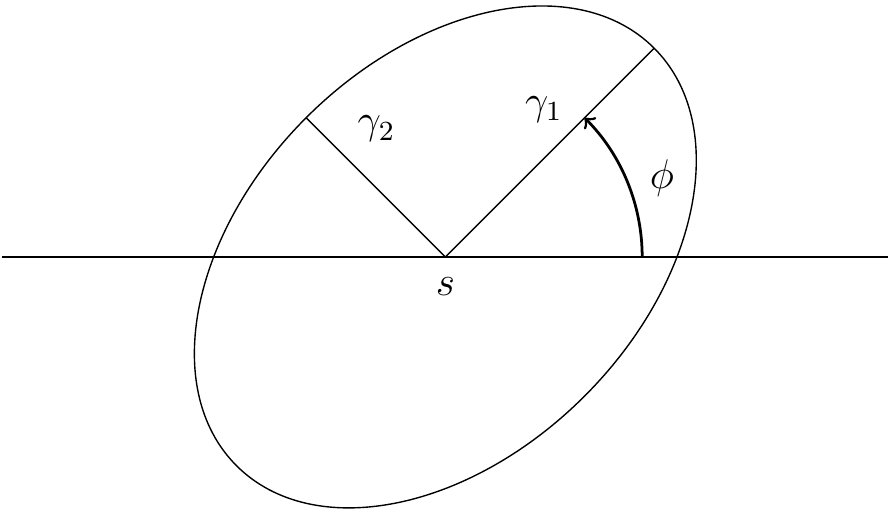}
			&
			\includegraphics[width=0.5\textwidth]{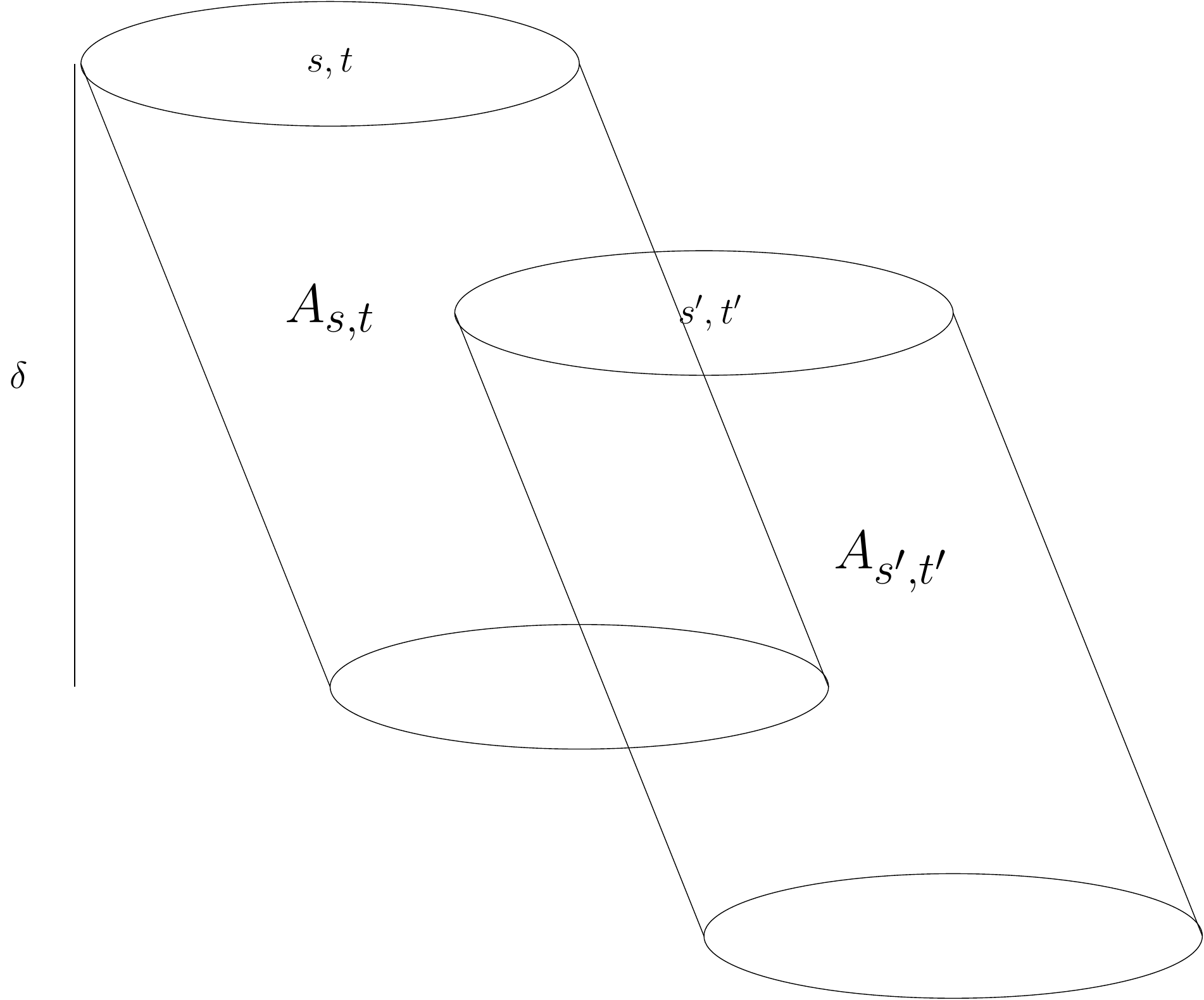}			
			\\
			(a) & (b)
		\end{tabular}
	\end{center} 
	\caption{Space-time kernels. Left display: a spatial ellipse $E(s,\gamma_1,\gamma_2,\phi)$
		centered at $s$. Right display: an intersection of two slated elliptical cylinders $A_{s,t}$ and $A_{s',t'}$ with duration $\delta$. }
	\label{fig:examples}
\end{figure}


\clearpage
\begin{figure}[th!]
	\begin{center}
		\begin{tabular}{ccc}
			\includegraphics[width=0.3\textwidth]{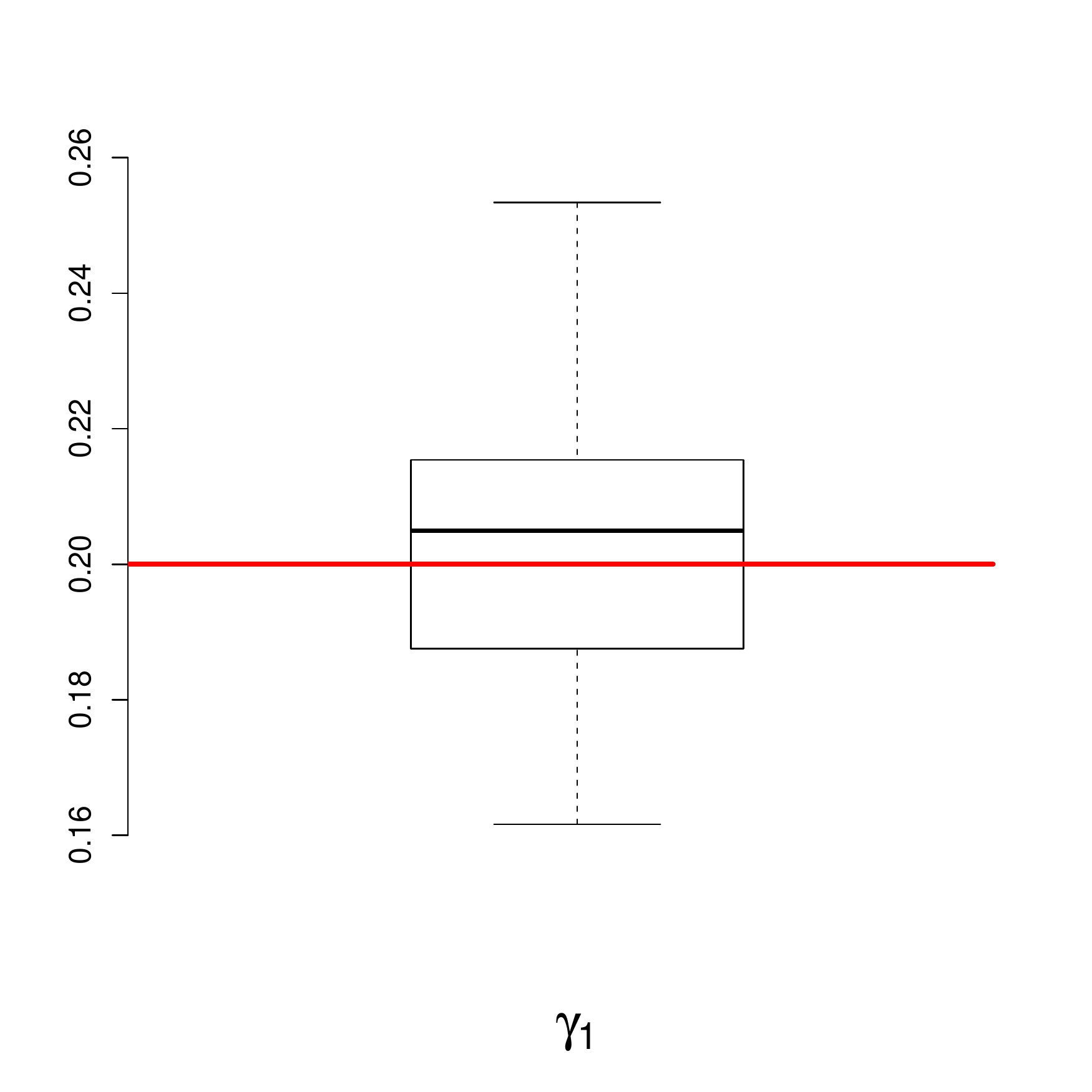}
			&
			\includegraphics[width=0.3\textwidth]{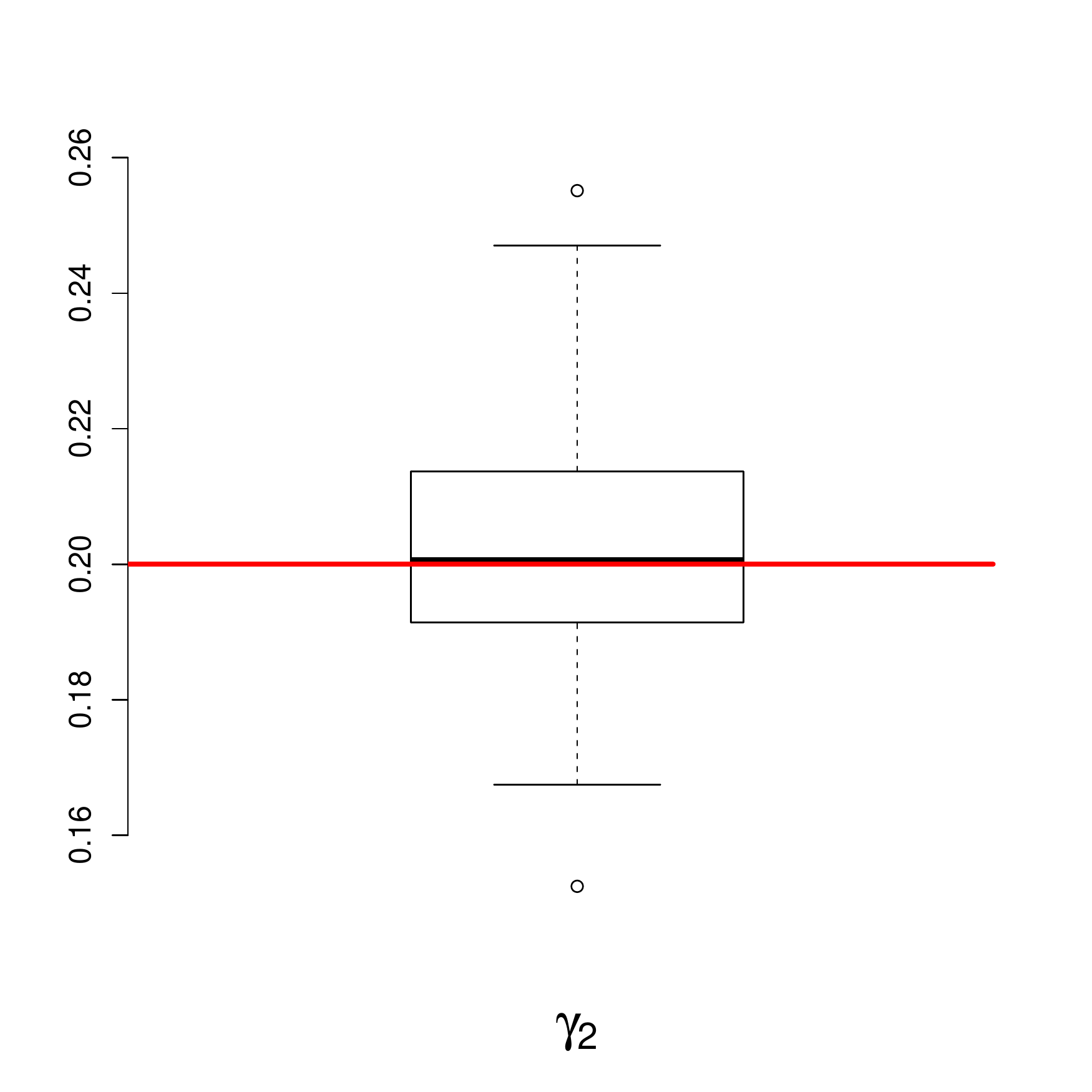}
			&
			\includegraphics[width=0.3\textwidth]{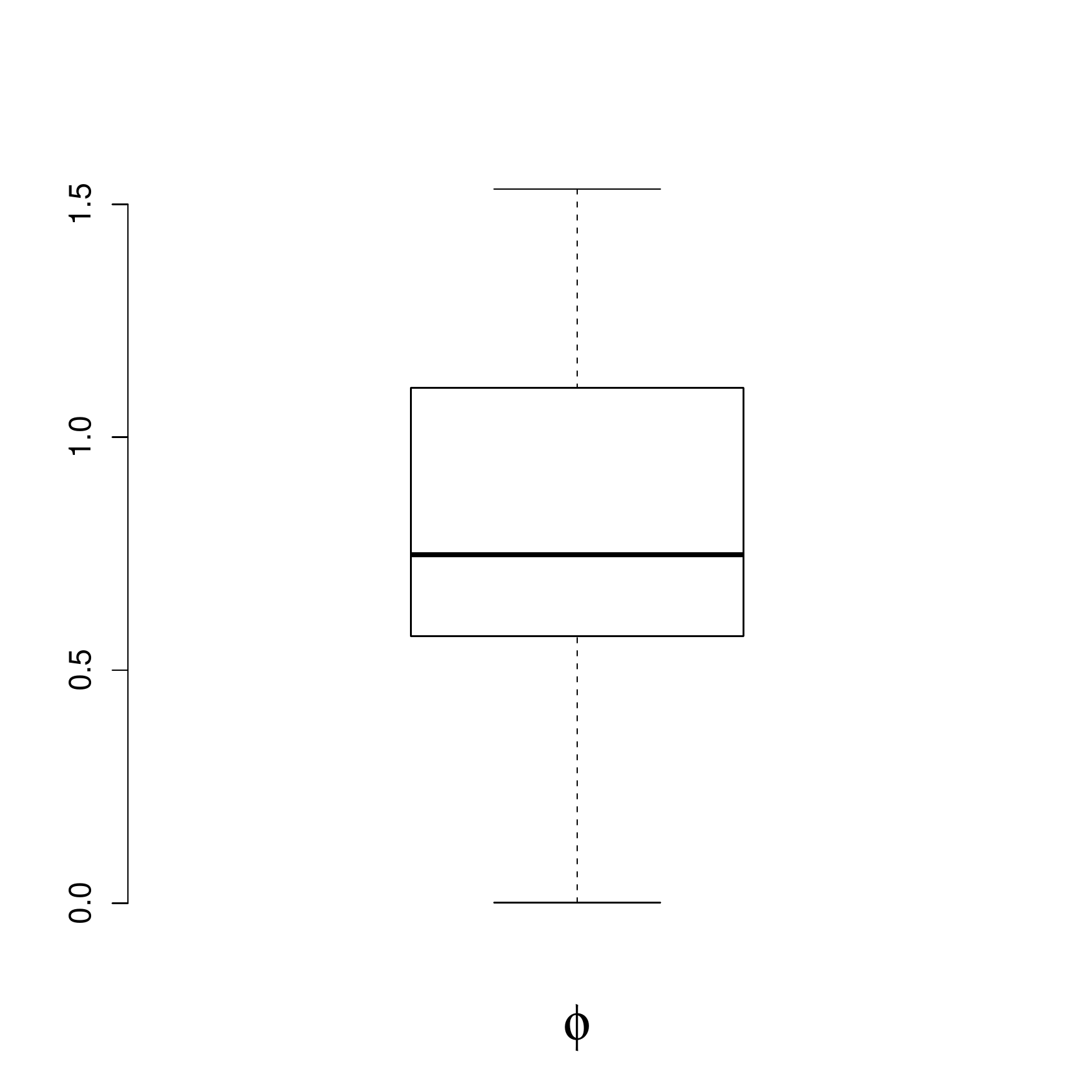}
			\\				
			\includegraphics[width=0.3\textwidth]{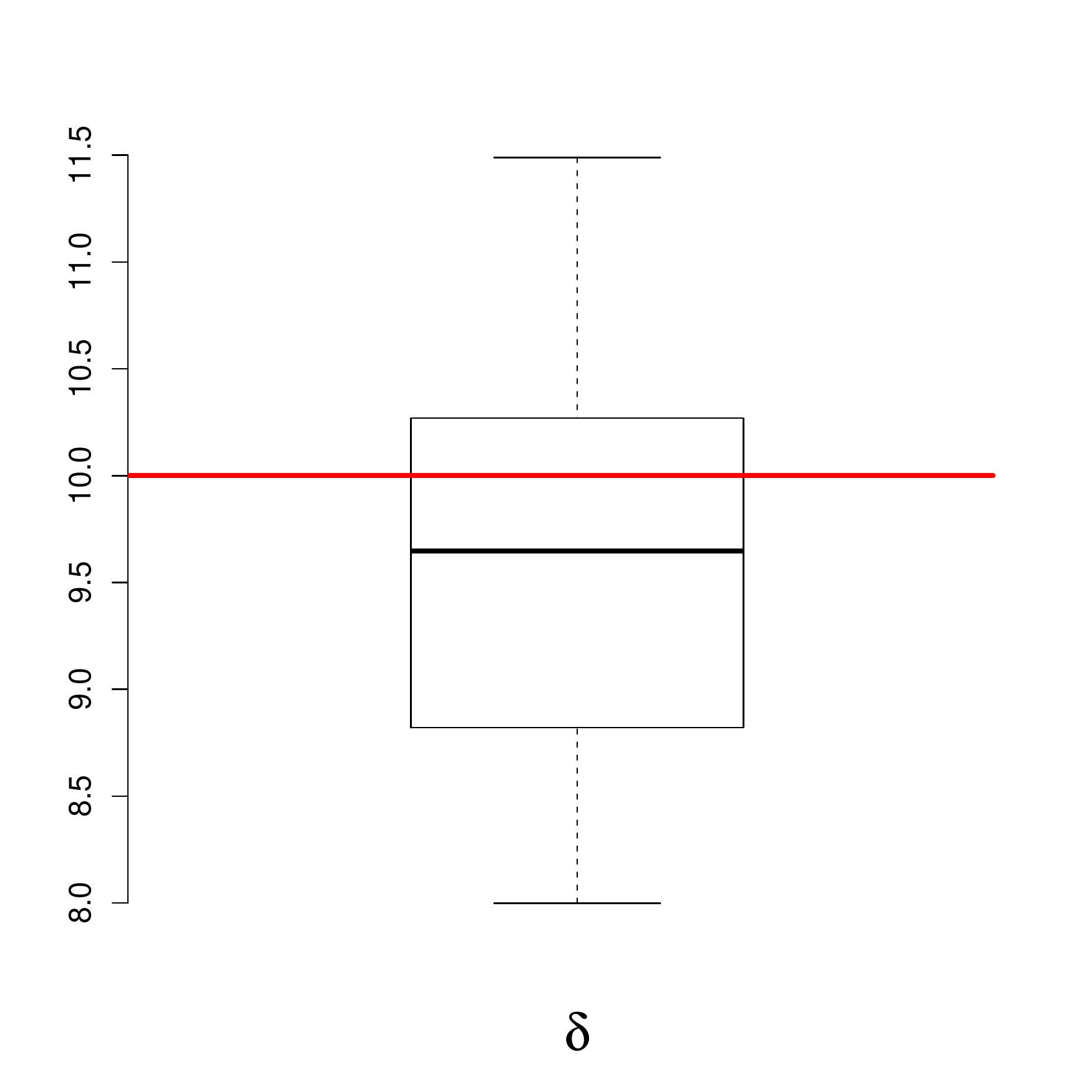}
			&
			\includegraphics[width=0.3\textwidth]{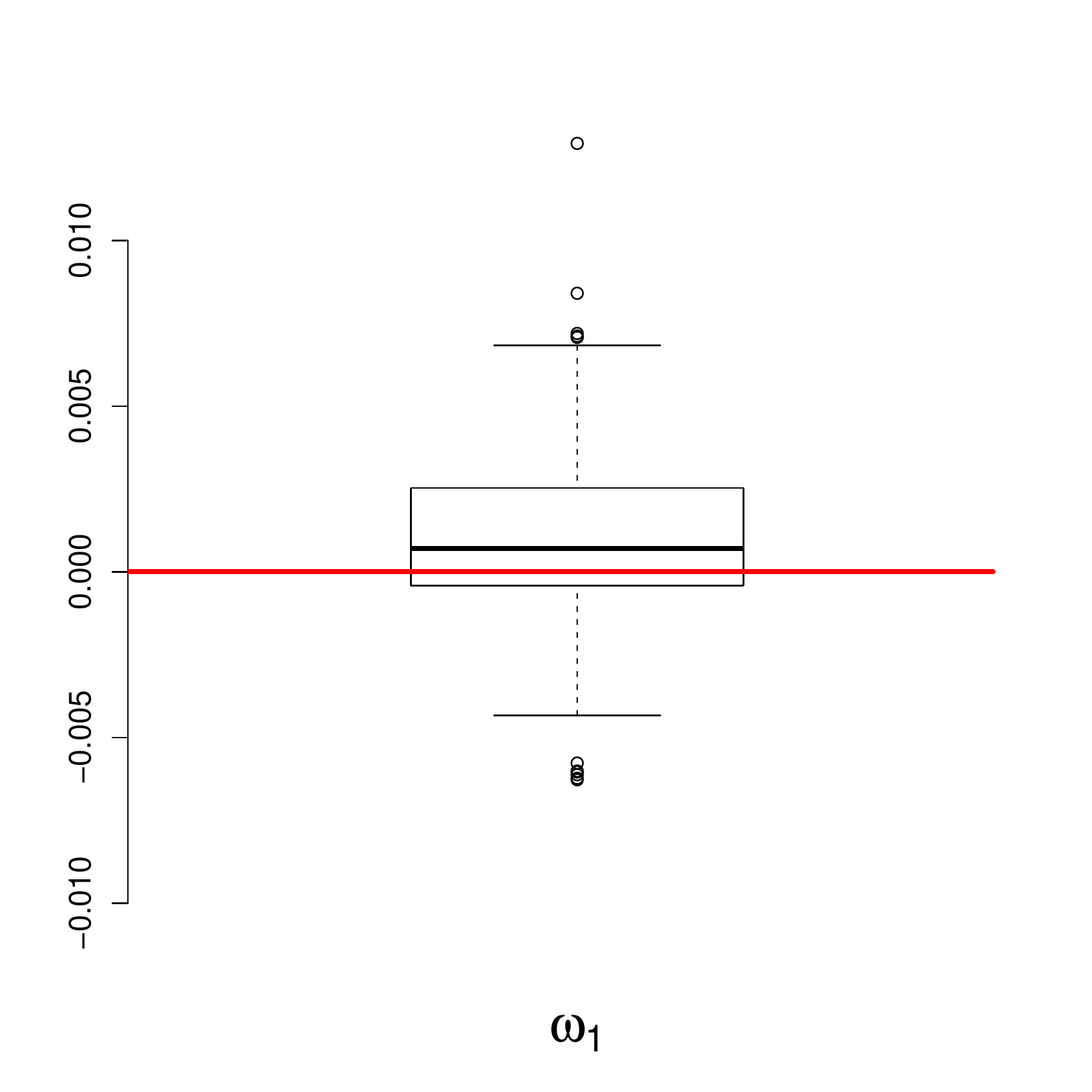}
			&
			\includegraphics[width=0.3\textwidth]{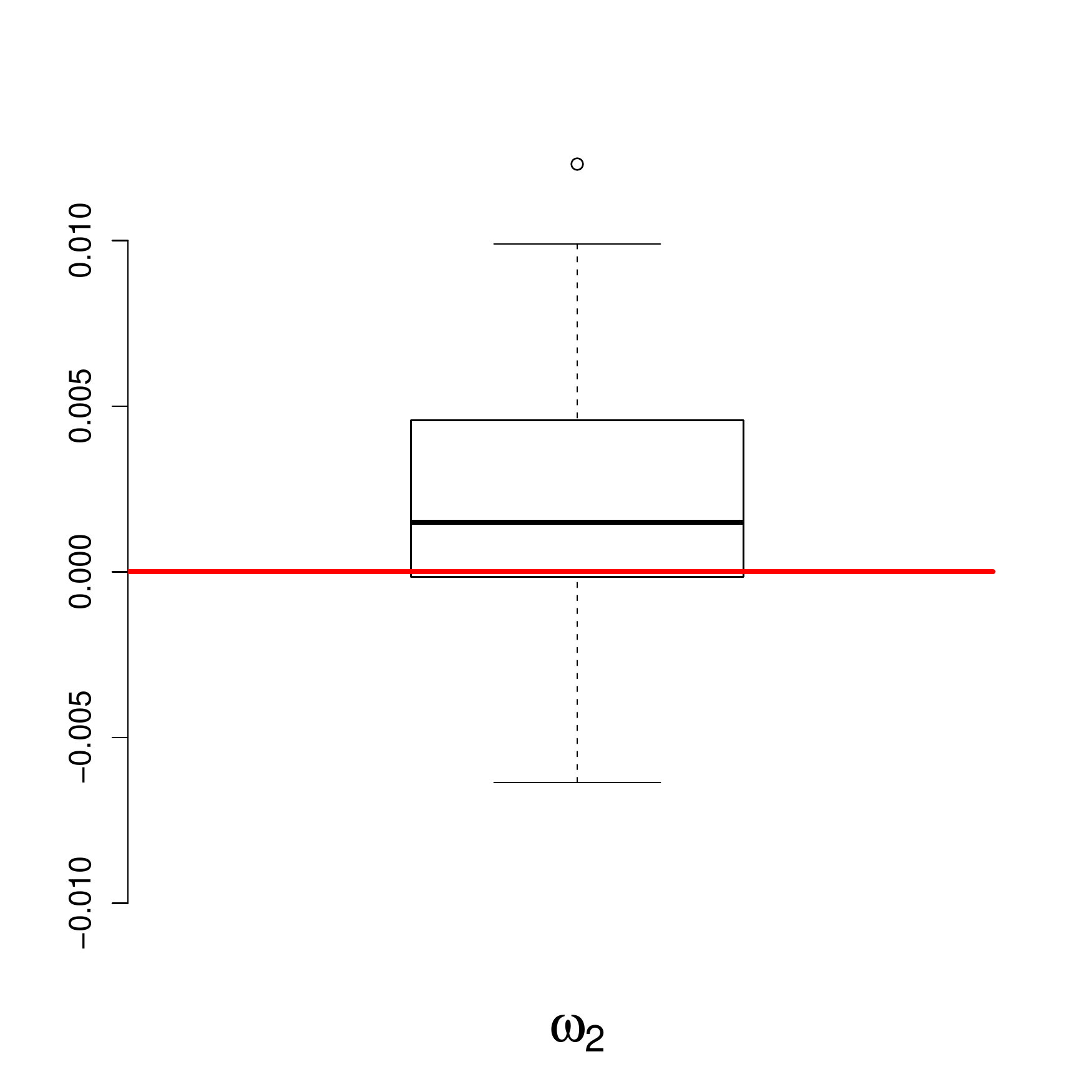}
		\end{tabular}
	\end{center} 
	\caption{Summary of parameter estimates for Scenario A of the simulation study: boxplots of parameter estimates for $100$ simulated datasets.}\label{fig:boxplotsA}
\end{figure} 

\clearpage

\begin{figure}[th!]
	\begin{center}
		\begin{tabular}{ccc}
			\includegraphics[width=0.3\textwidth]{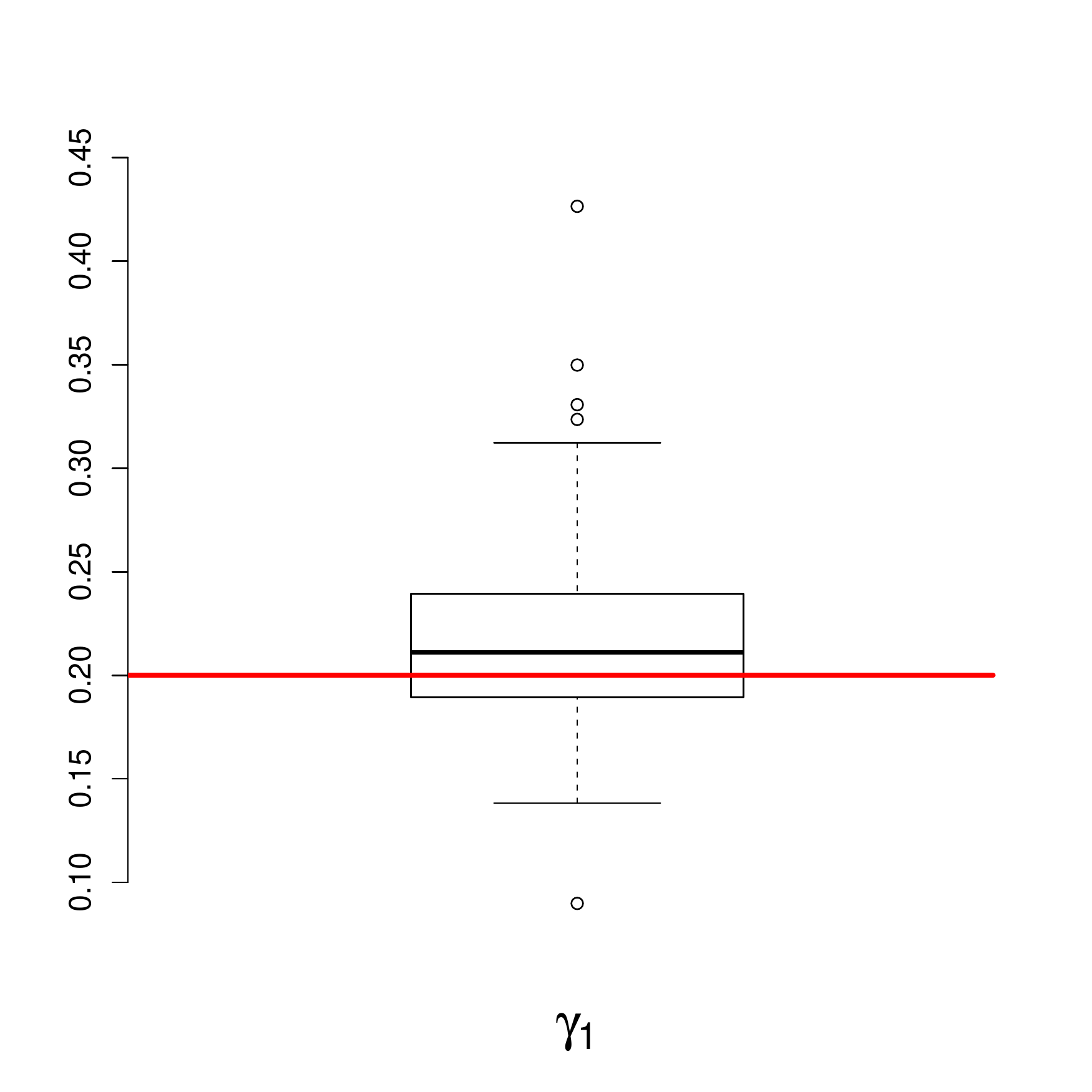}
			&
			\includegraphics[width=0.3\textwidth]{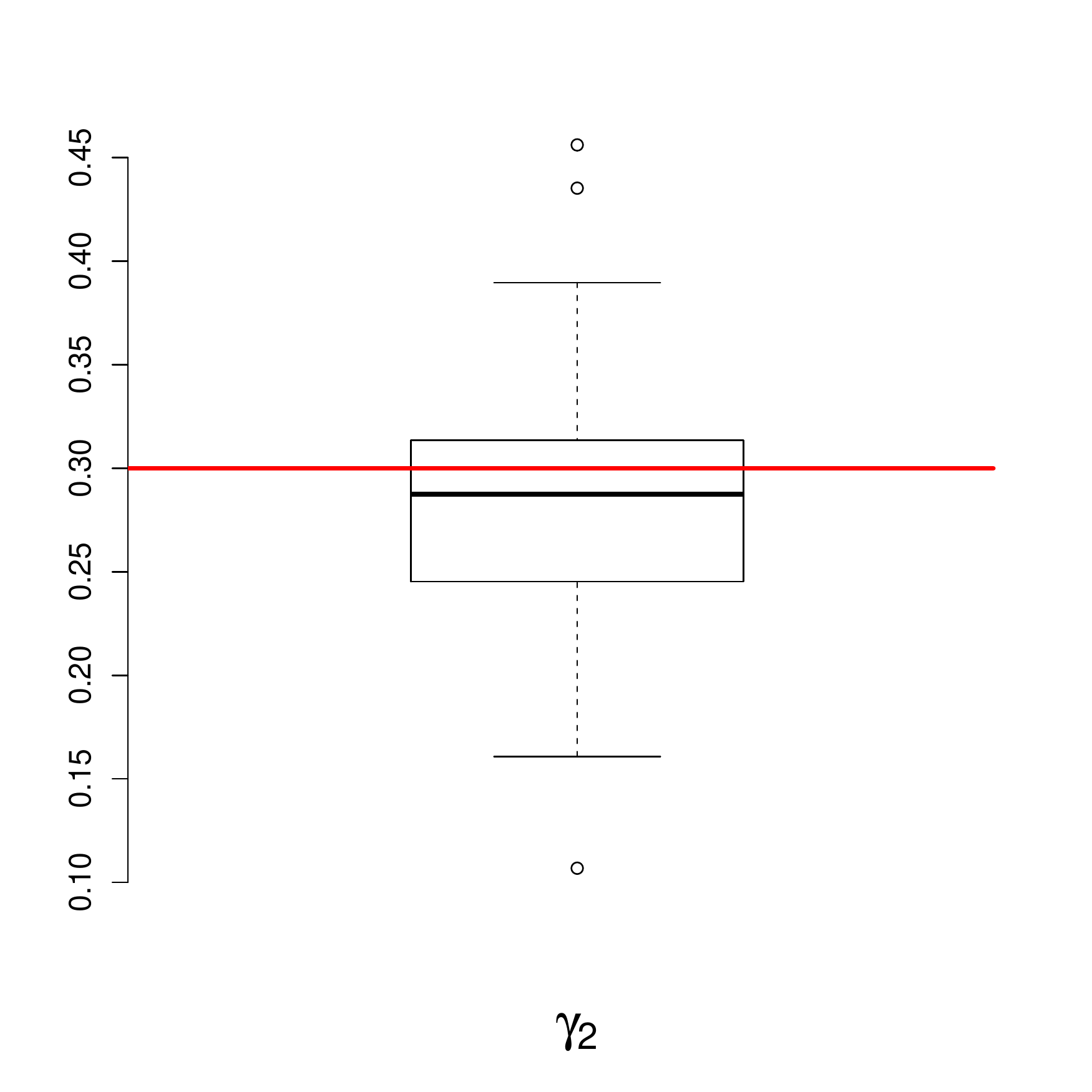}
			&
			\includegraphics[width=0.3\textwidth]{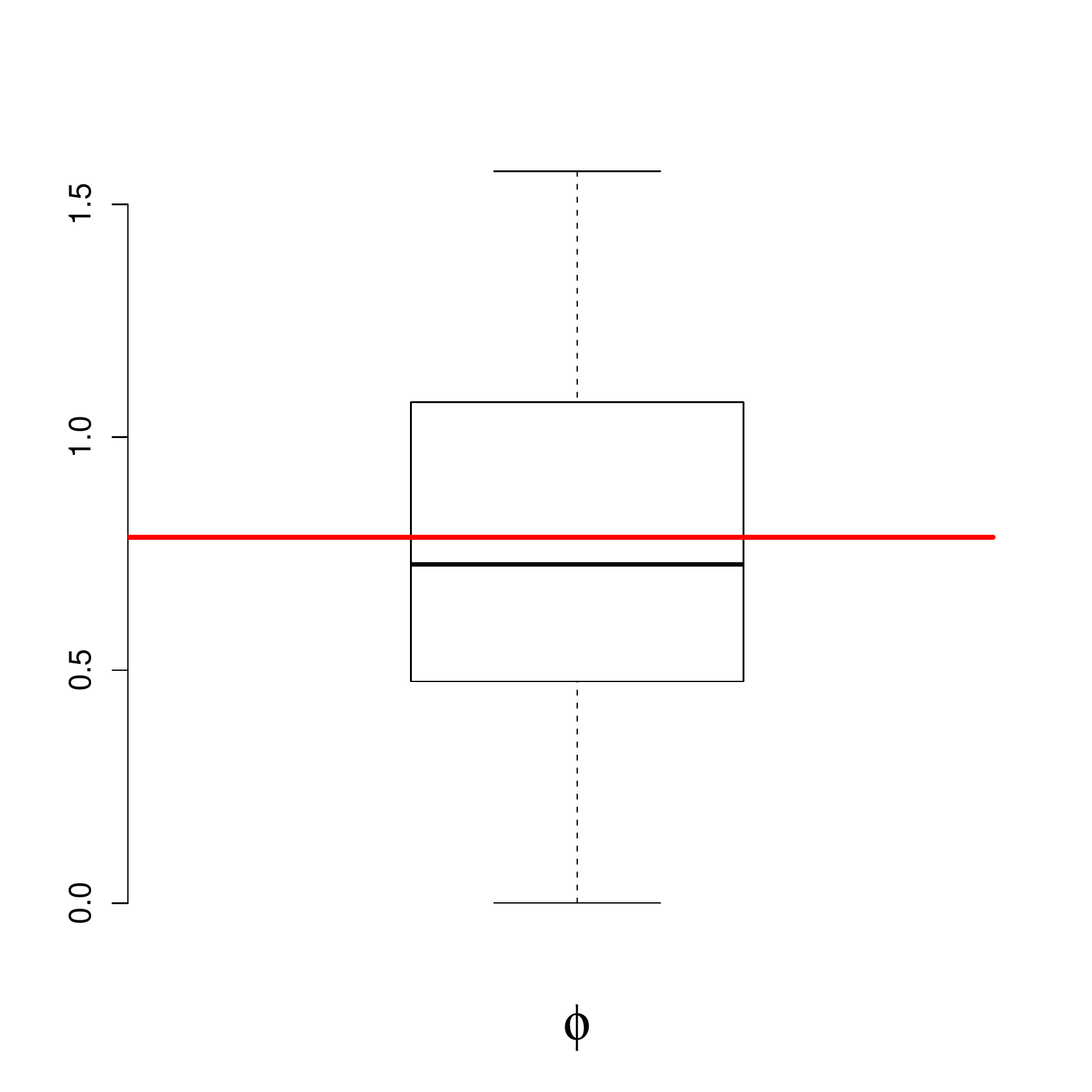}
			\\				
			\includegraphics[width=0.3\textwidth]{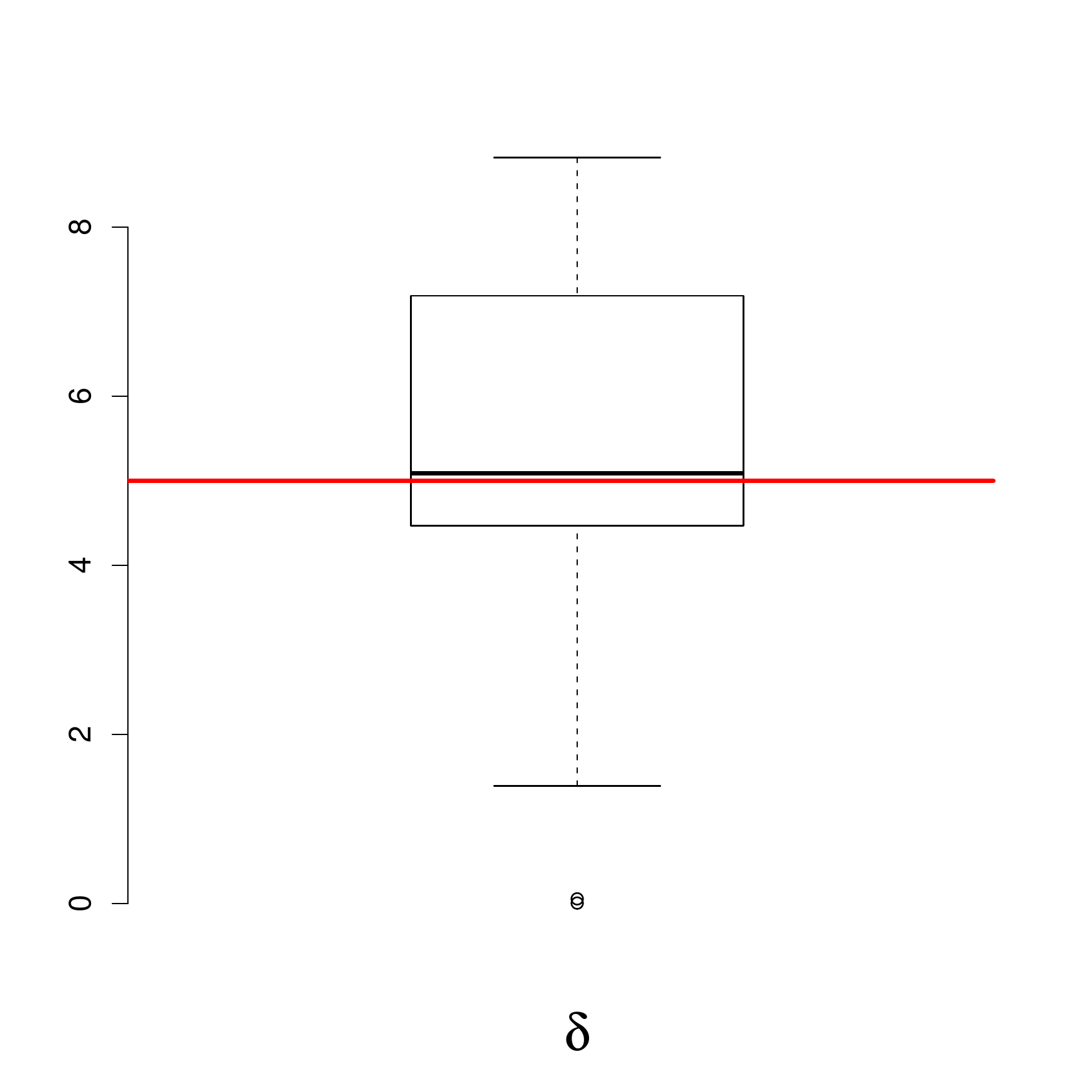}
			&
			\includegraphics[width=0.3\textwidth]{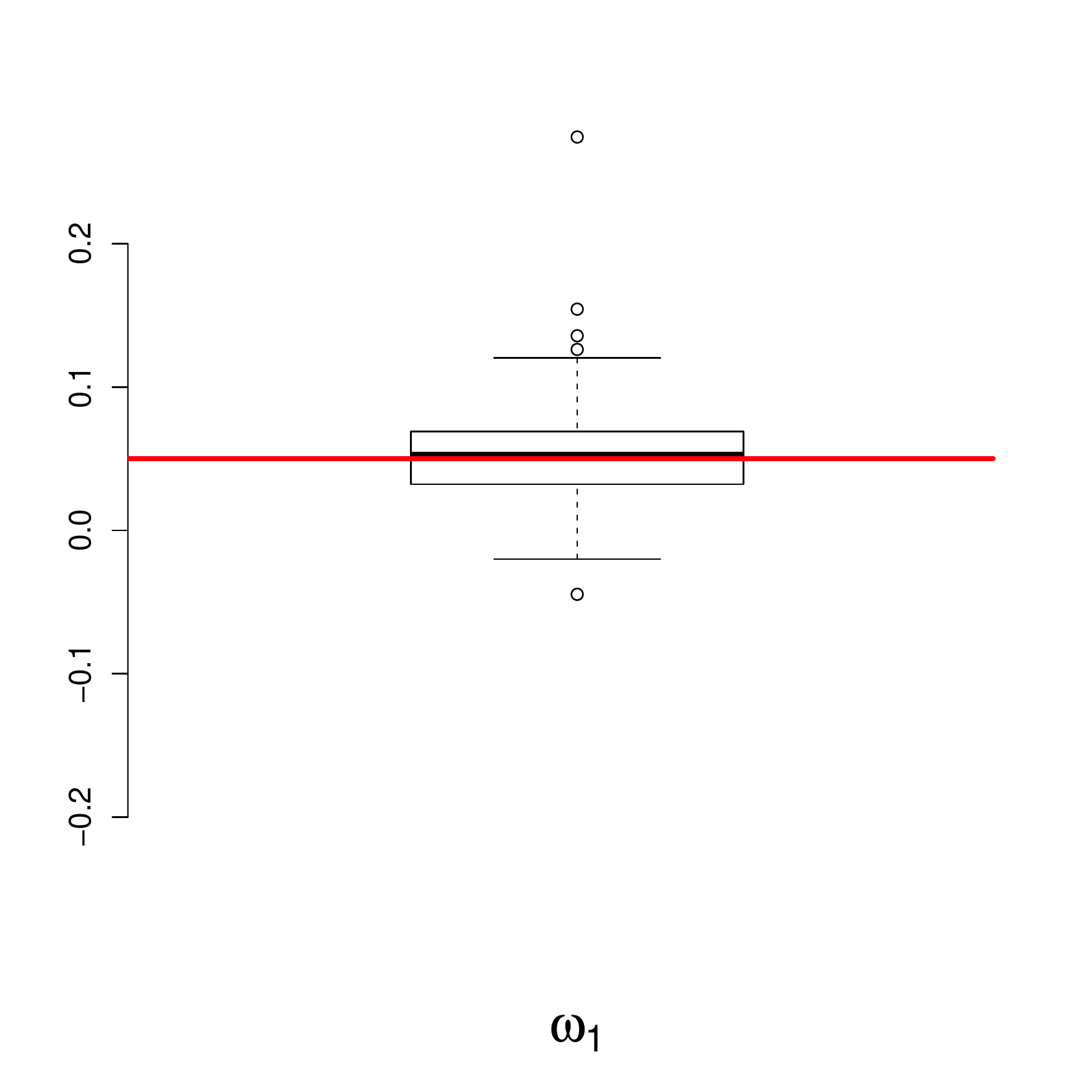}
			&
			\includegraphics[width=0.3\textwidth]{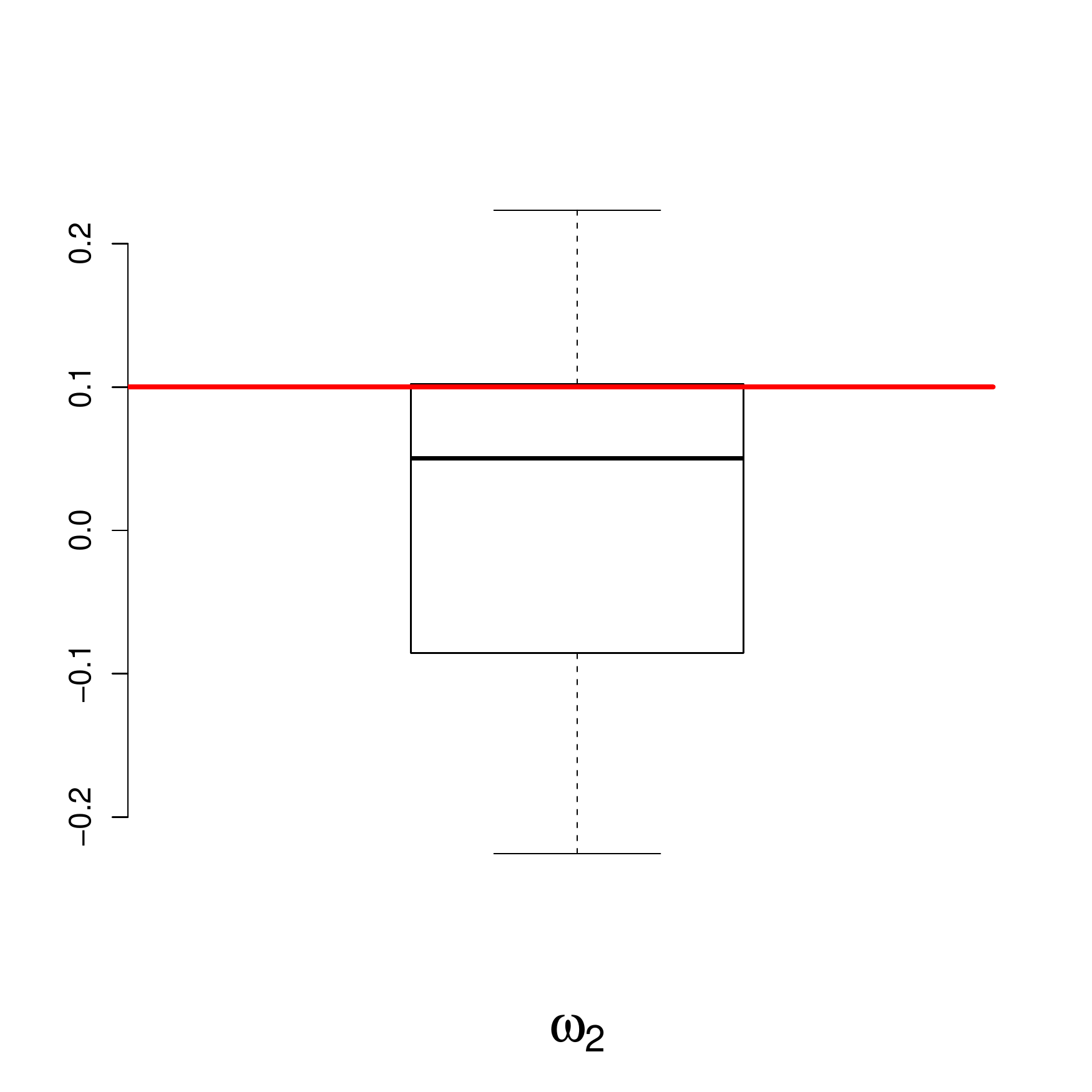}
		\end{tabular}
	\end{center} 
	\caption{Summary of parameter estimates for Scenario B of the simulation study: boxplots of parameter estimates for $100$  simulated datasets.}\label{fig:boxplotsD}
\end{figure} 

\clearpage
\begin{figure}[th!]
	\begin{center}
		\includegraphics[width=0.6\textwidth]{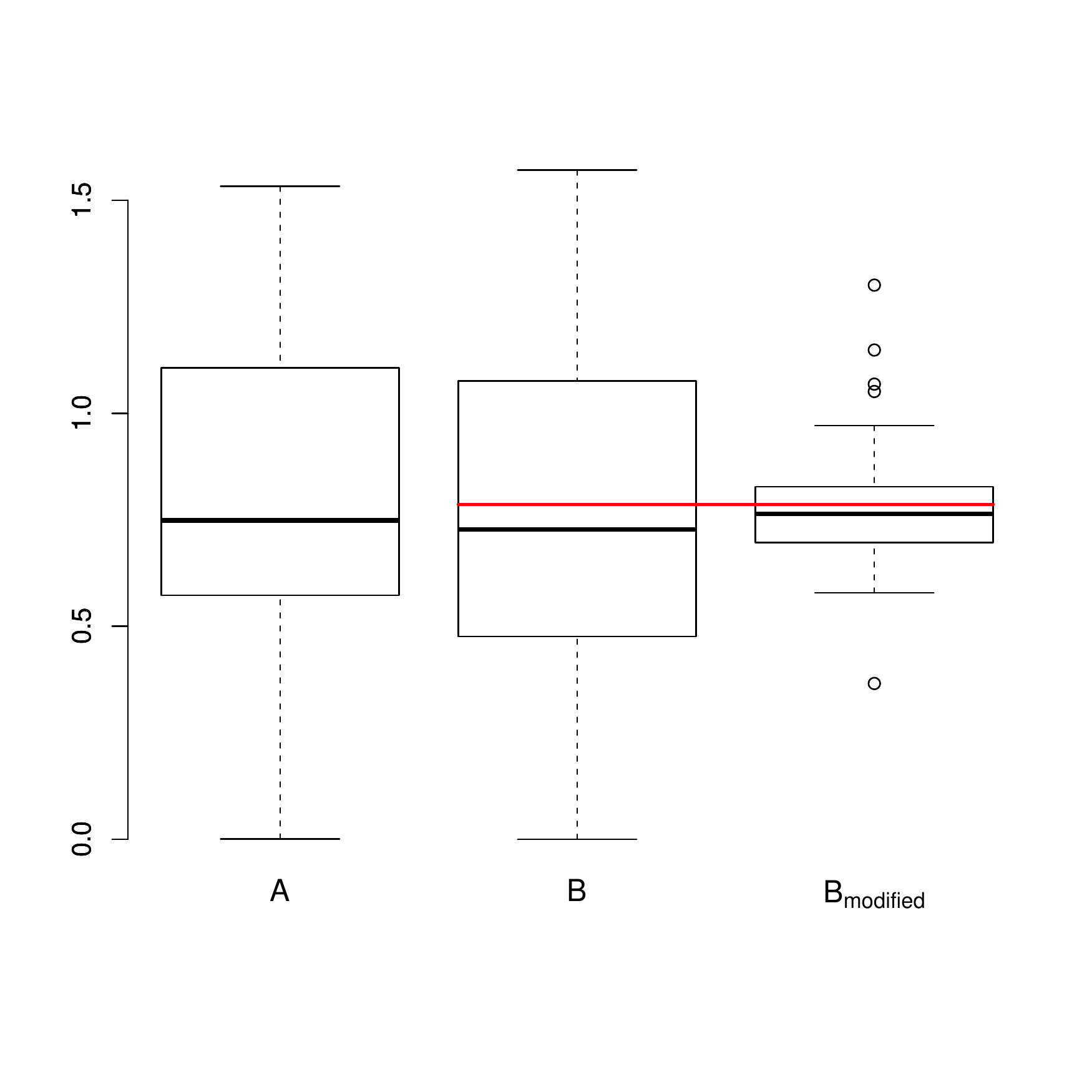}
	\end{center} 
	\caption{Parameter estimates of the simulation study: boxplots of $\phi$ estimates according to scenario A, scenario B and a modified scenario B  with $\gamma_2=0.5$.}\label{fig:phi}
\end{figure} 

\clearpage
\begin{figure}[t!]
	\begin{center}
		\begin{tabular}{cc}
								\includegraphics[width=0.38\textwidth,height=0.34\textheight]{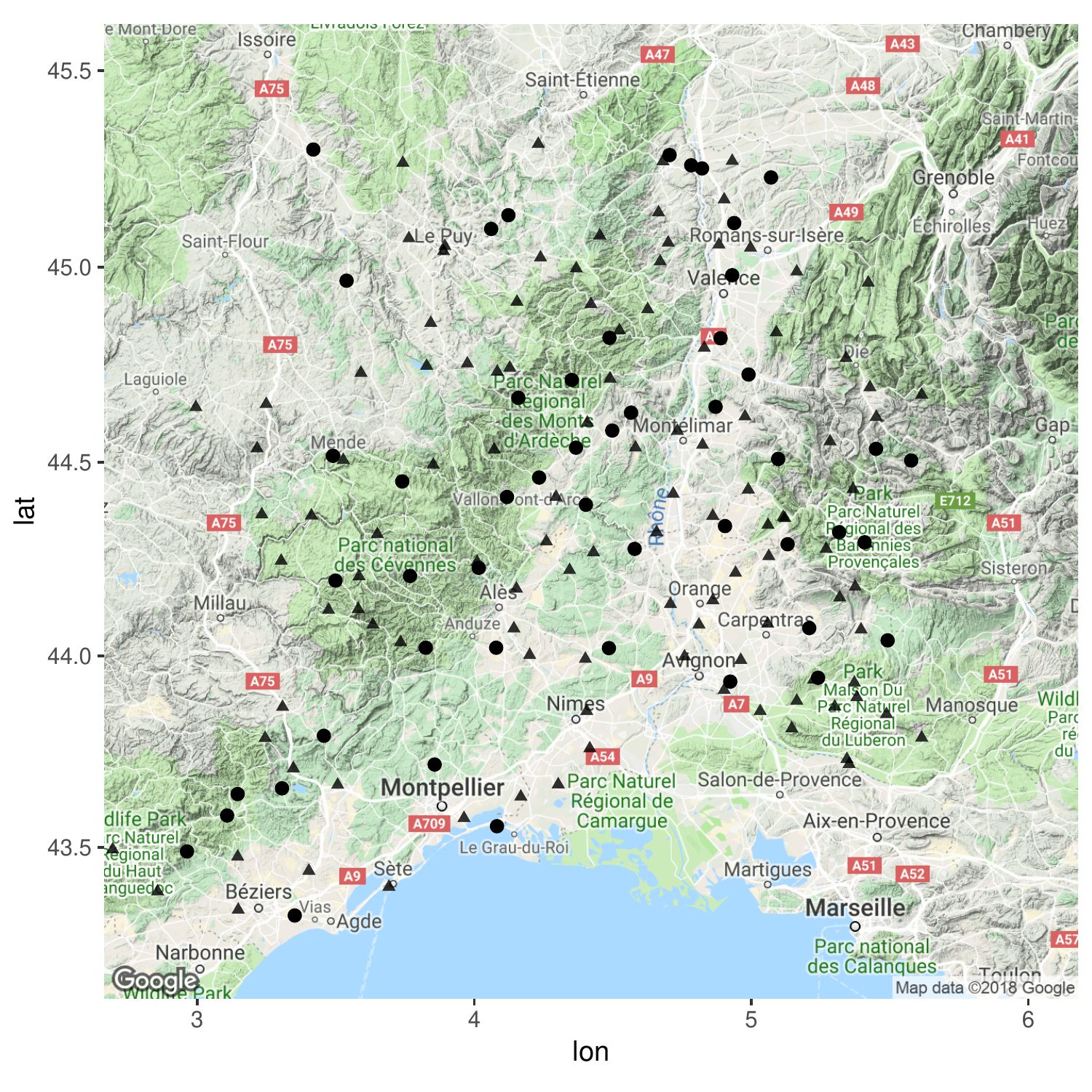} &
				\includegraphics[width=0.4\textwidth]{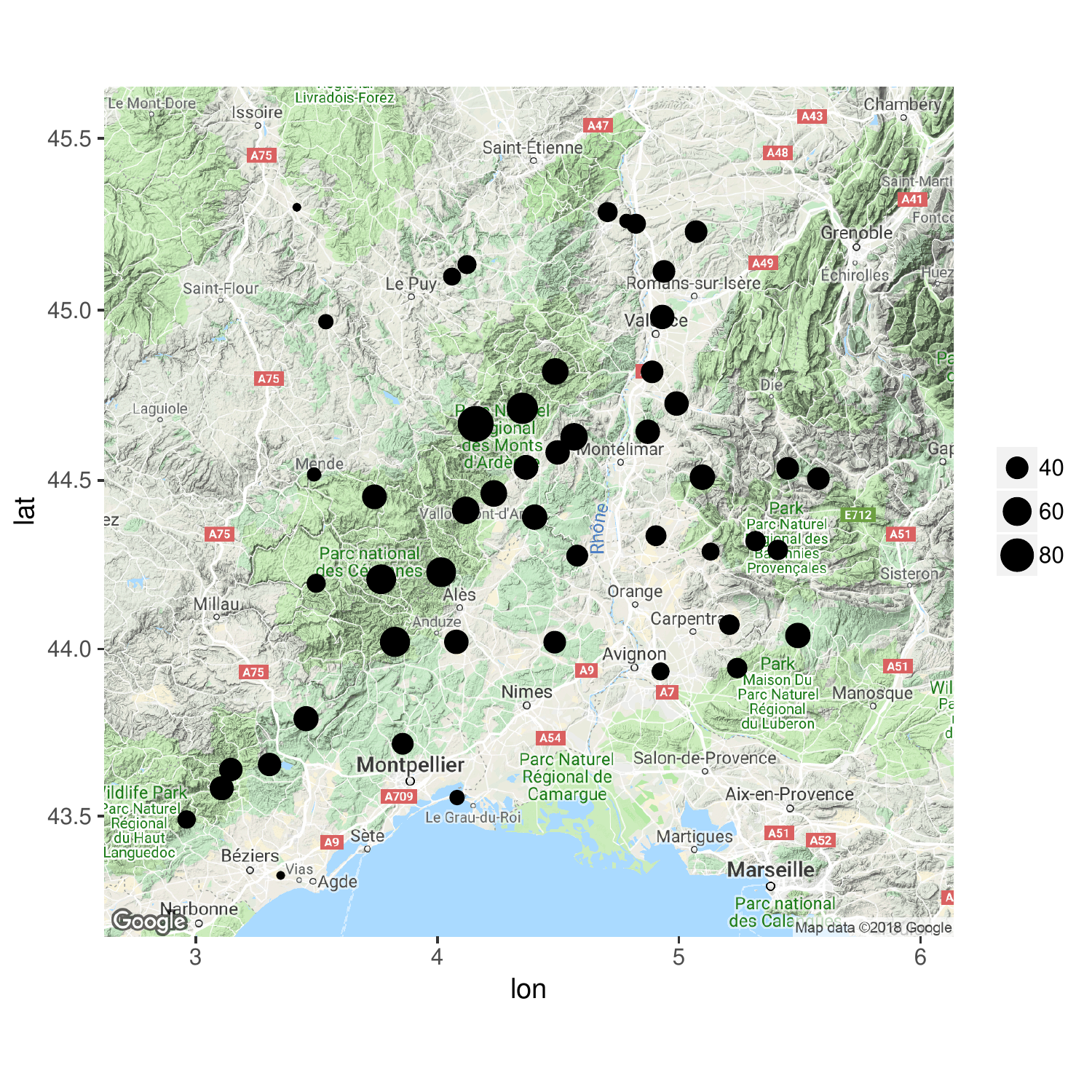}
\\	
			&$u$ \\
			\includegraphics[width=0.4\textwidth]{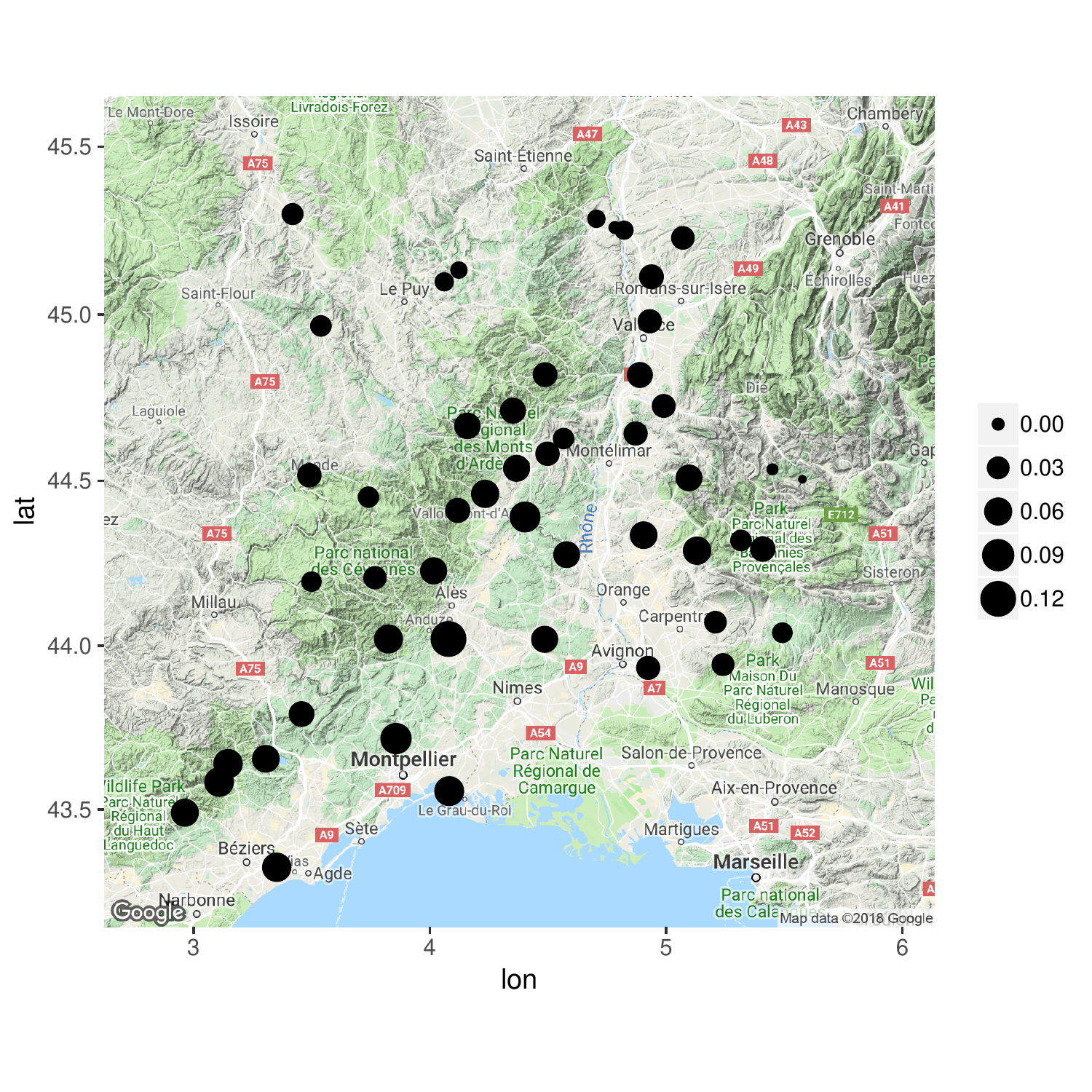}
			&
			\includegraphics[width=0.4\textwidth]{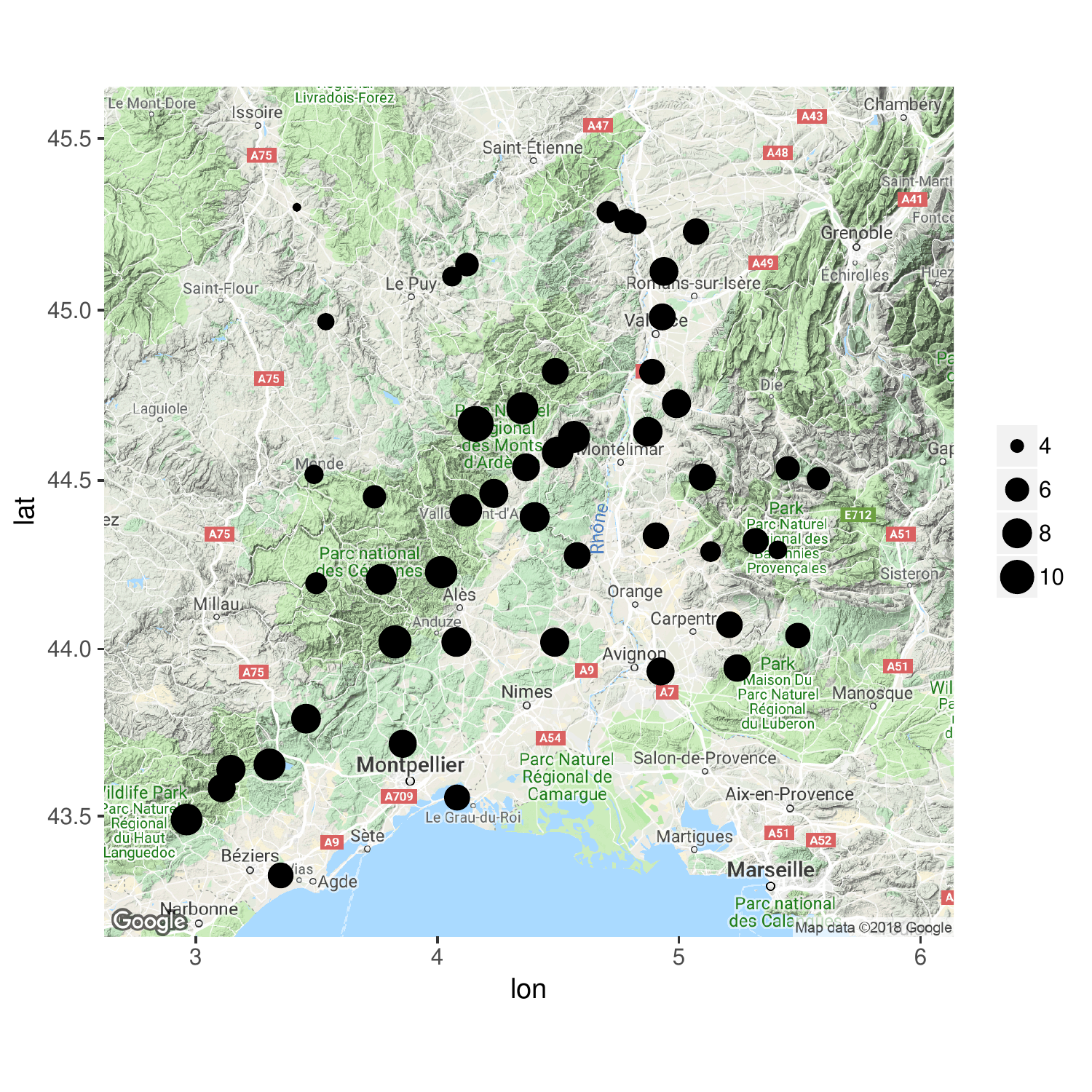}
			\\
			$\xi$ &$\sigma$
		\end{tabular}		
		\caption{\textit{Precipitation data of Southern France. Top left display: topographic map showing the meteorological stations selected for our case study. Dots correspond to the stations used for fitting. In the other displays, their diameter is  proportional to  empirical \textit{$99\%$ quantiles} $u(s)$ (top right plot) and  to estimates of the GPD parameters  $\xi(s)$ (bottom left plot) and $\sigma(s)$ (bottom right plot).}}\label{fig:map}
	\end{center} 
\end{figure} 

\clearpage
\begin{figure}[th!]
	\begin{center}
\includegraphics[width=.95\textwidth]{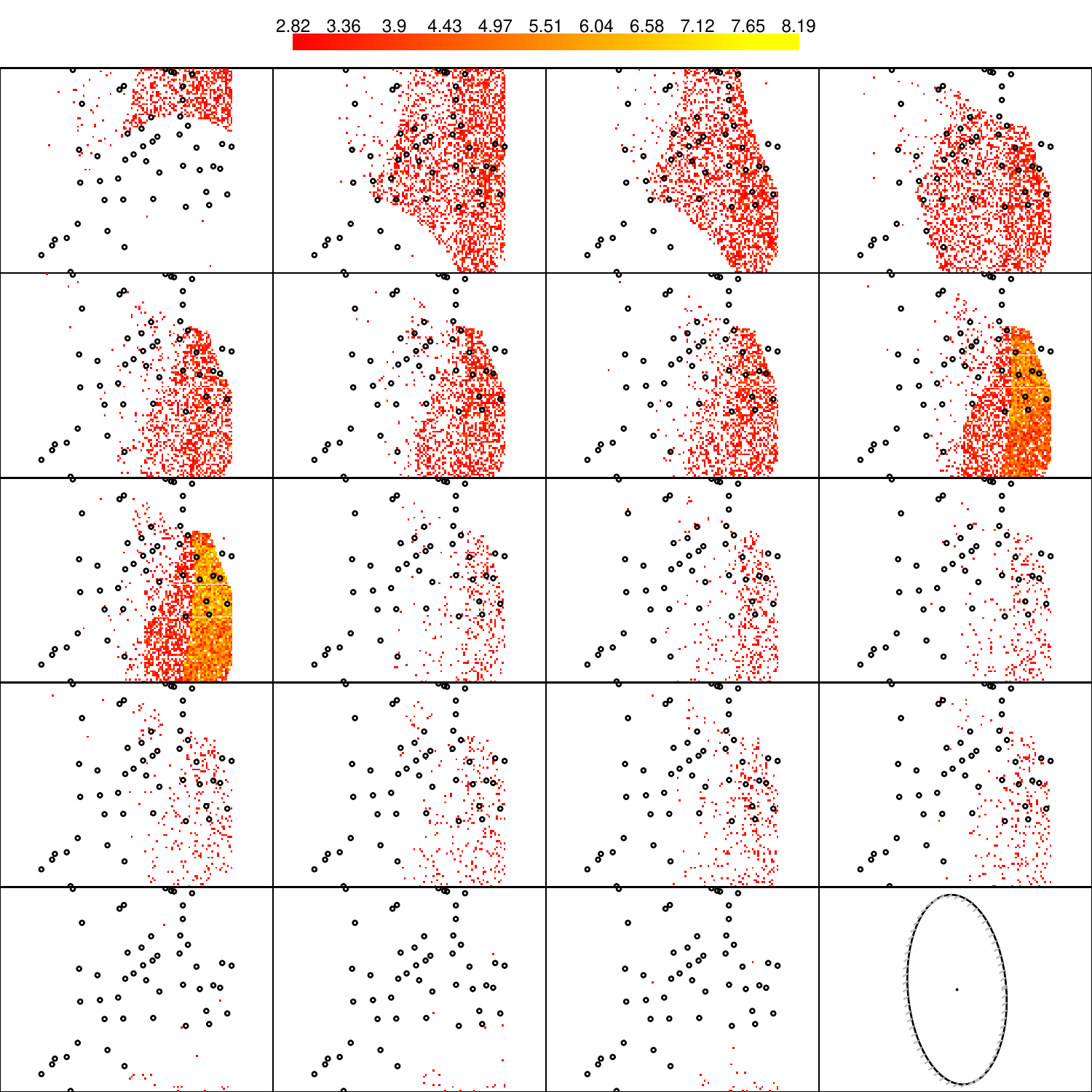}		
	\end{center} 
	\caption{\textit{A simulation example showing exceedances of the $0.95$-quantile for the  model G1 fitted to precipitation data. Dots correspond to the stations used for fitting. The evolution over time  during $19$ hours is presented row-wise starting from the top left. The bottom right display illustrates the 
 estimated ellipses, centred at the barycenter of the locations,  and the movement induced by the velocity vector.}} \label{fig:meteo-gamma}
\end{figure} 
\clearpage
%

\begin{figure}[th!]
    \begin{center}
        \begin{tabular}{cc}
            \includegraphics[width=0.4\textwidth]{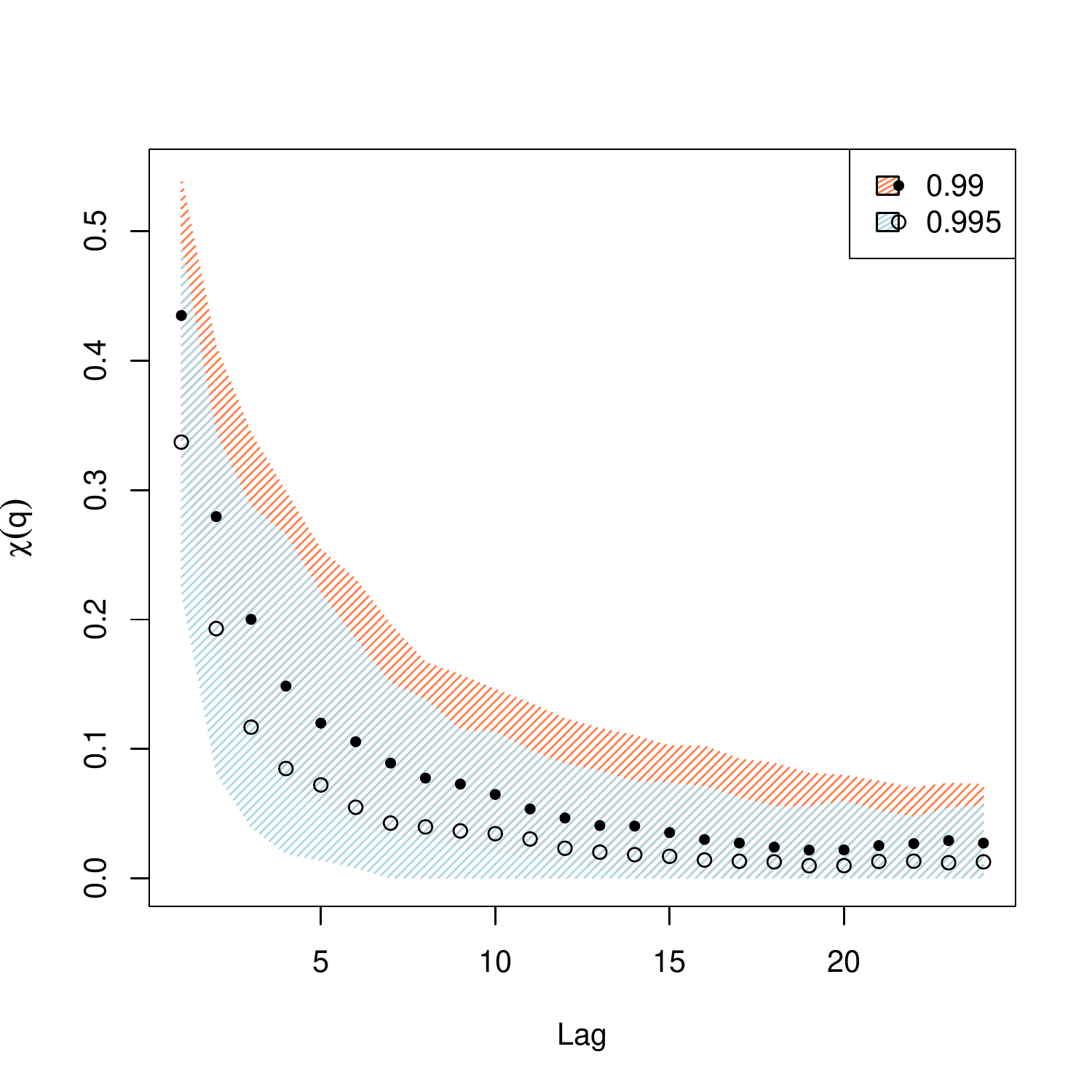}
            &
            \includegraphics[width=0.4\textwidth]{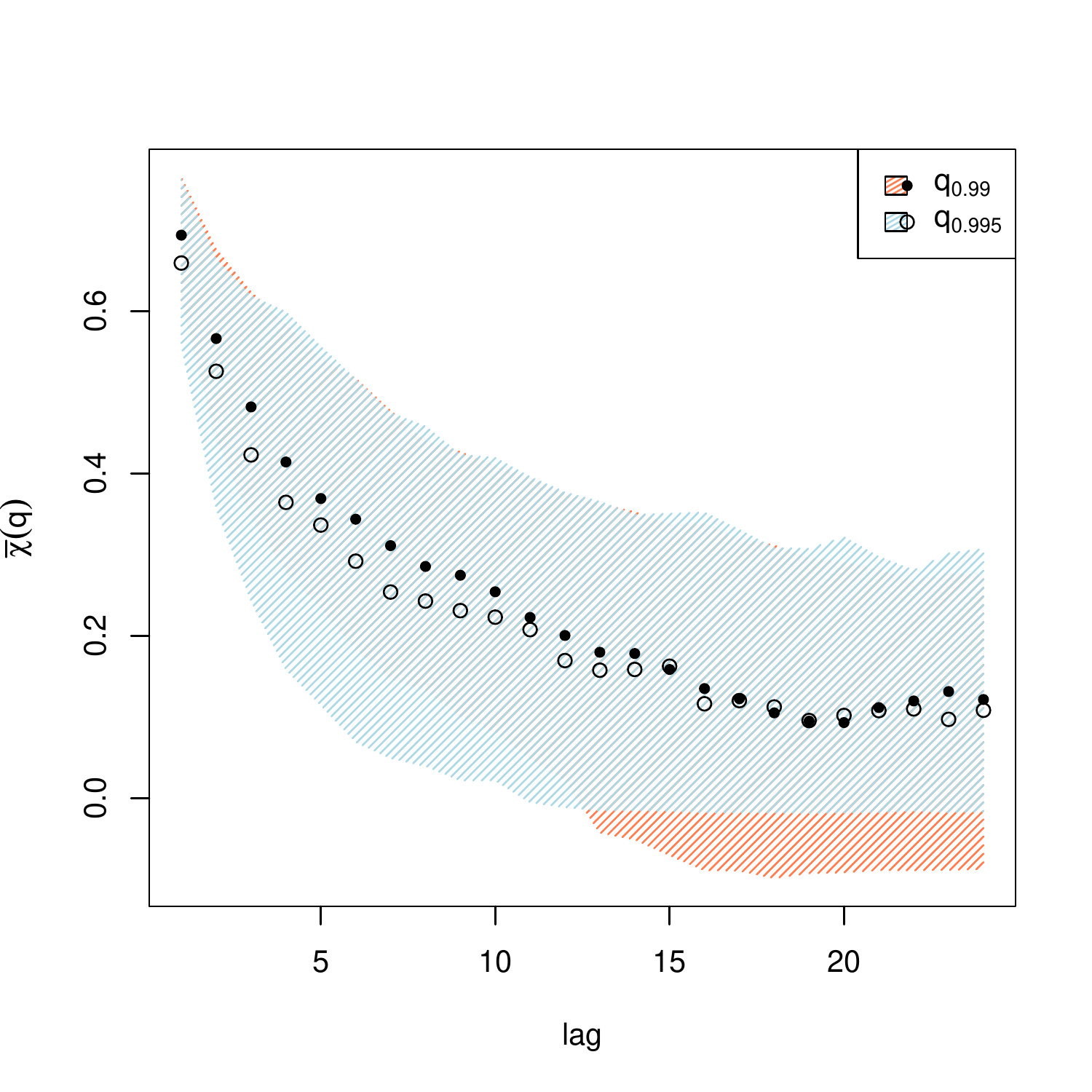}
            \\
\includegraphics[width=0.4\textwidth]{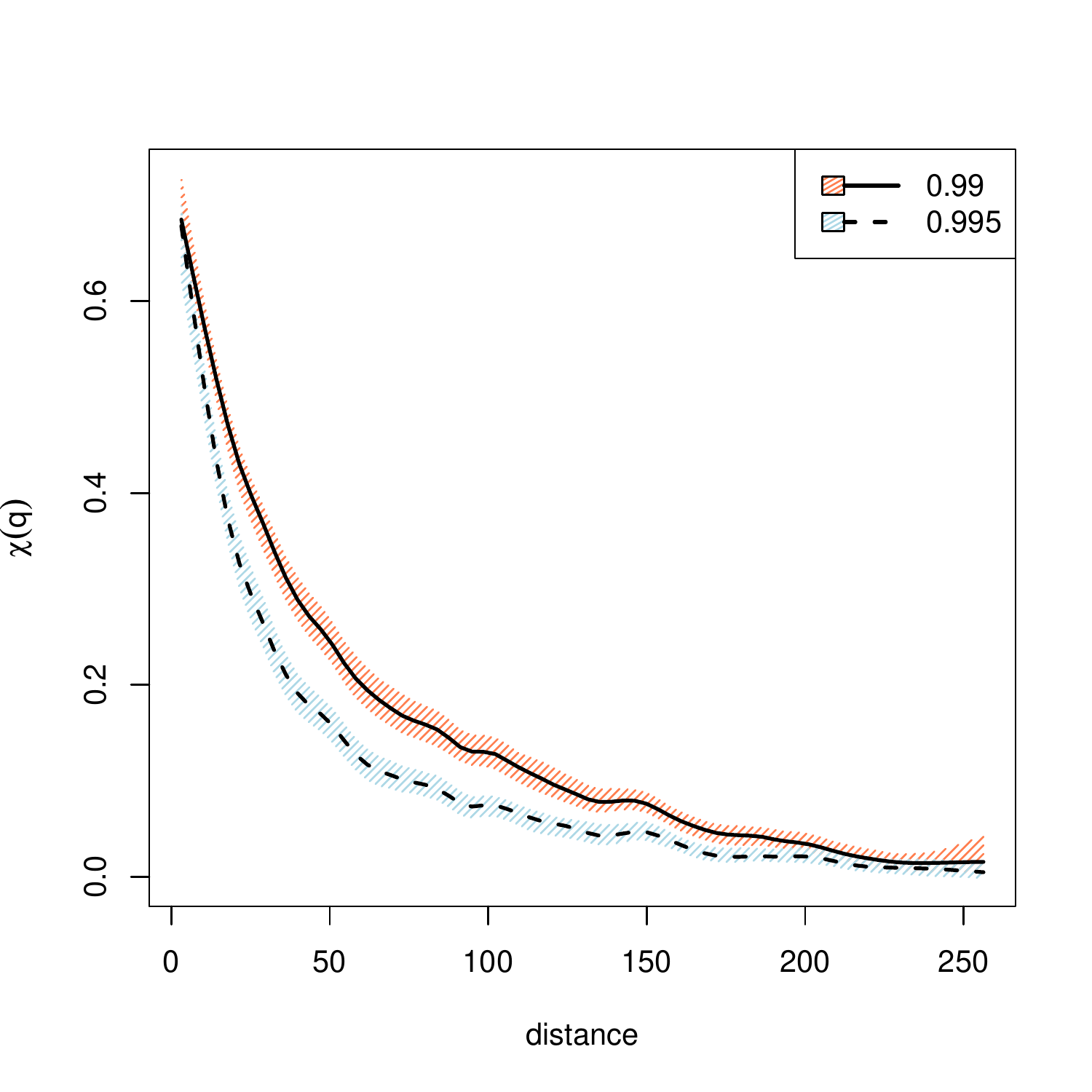}
            &
\includegraphics[width=0.4\textwidth]{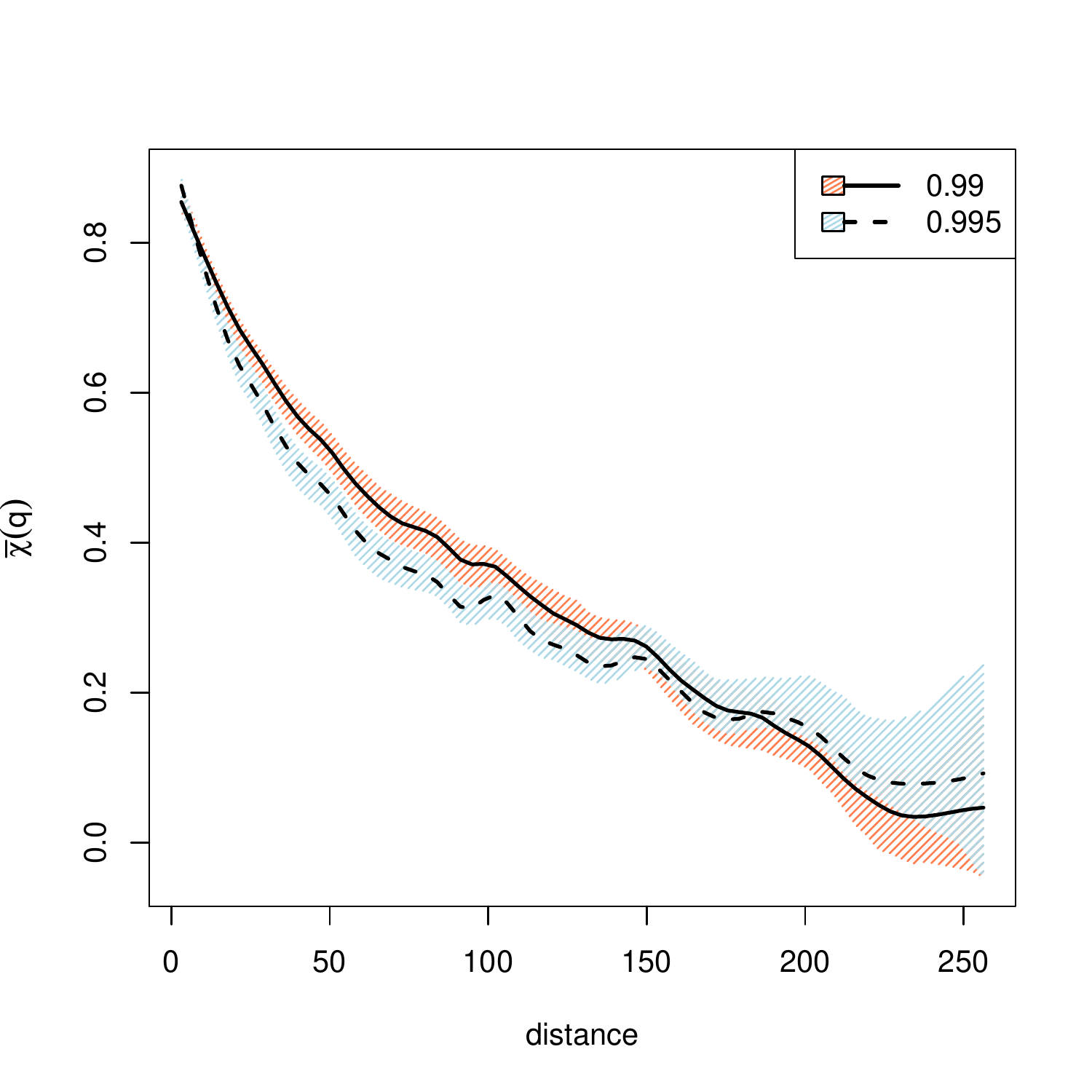}
        \end{tabular}

    \end{center}

    \caption{
        Empirical estimates  of ${\chi}_x(q)$
        (left panels) and ${\bar\chi}_x(q)$ (right panels) coefficients for the precipitation data.
        The filled region represents an approximate 95\% confidence region based on a stationary bootstrap procedure. 
    } \label{fig:empchi}
\end{figure}

\clearpage
 \begin{figure}[th!]
	\begin{tabular}{ccc}
		\includegraphics[width=0.3\linewidth]{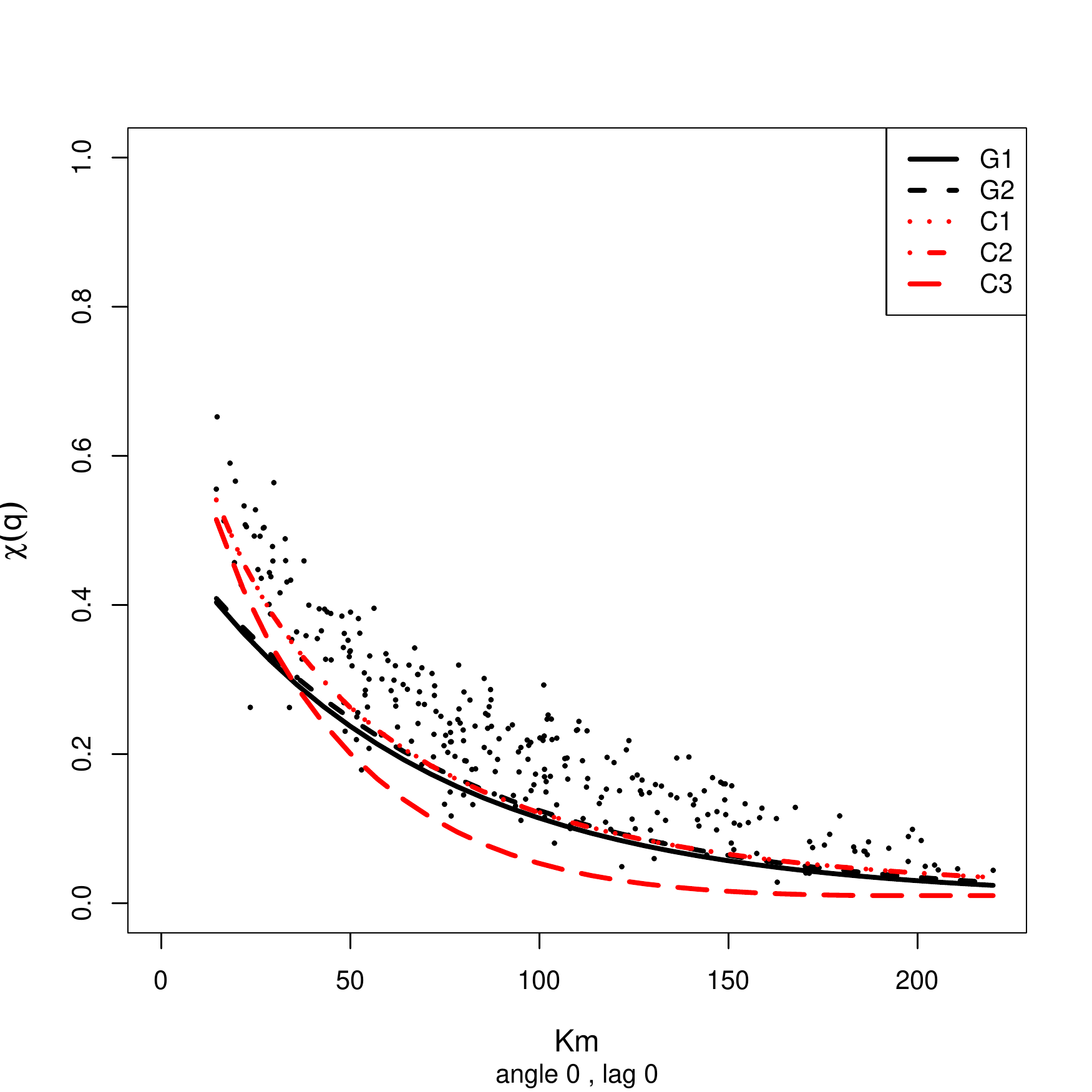}
		&
		\includegraphics[width=0.3\linewidth]{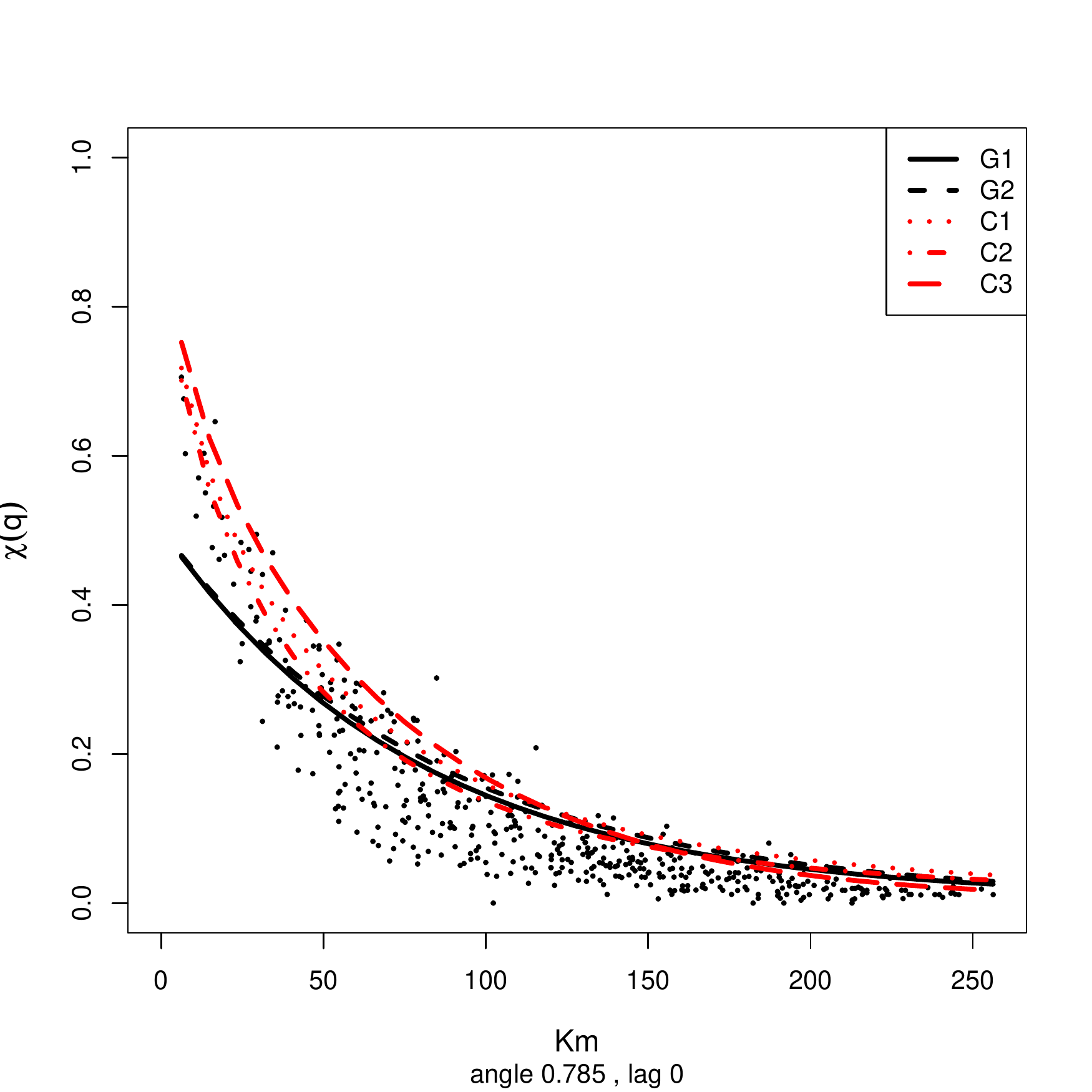}
		&
		\includegraphics[width=0.3\linewidth]{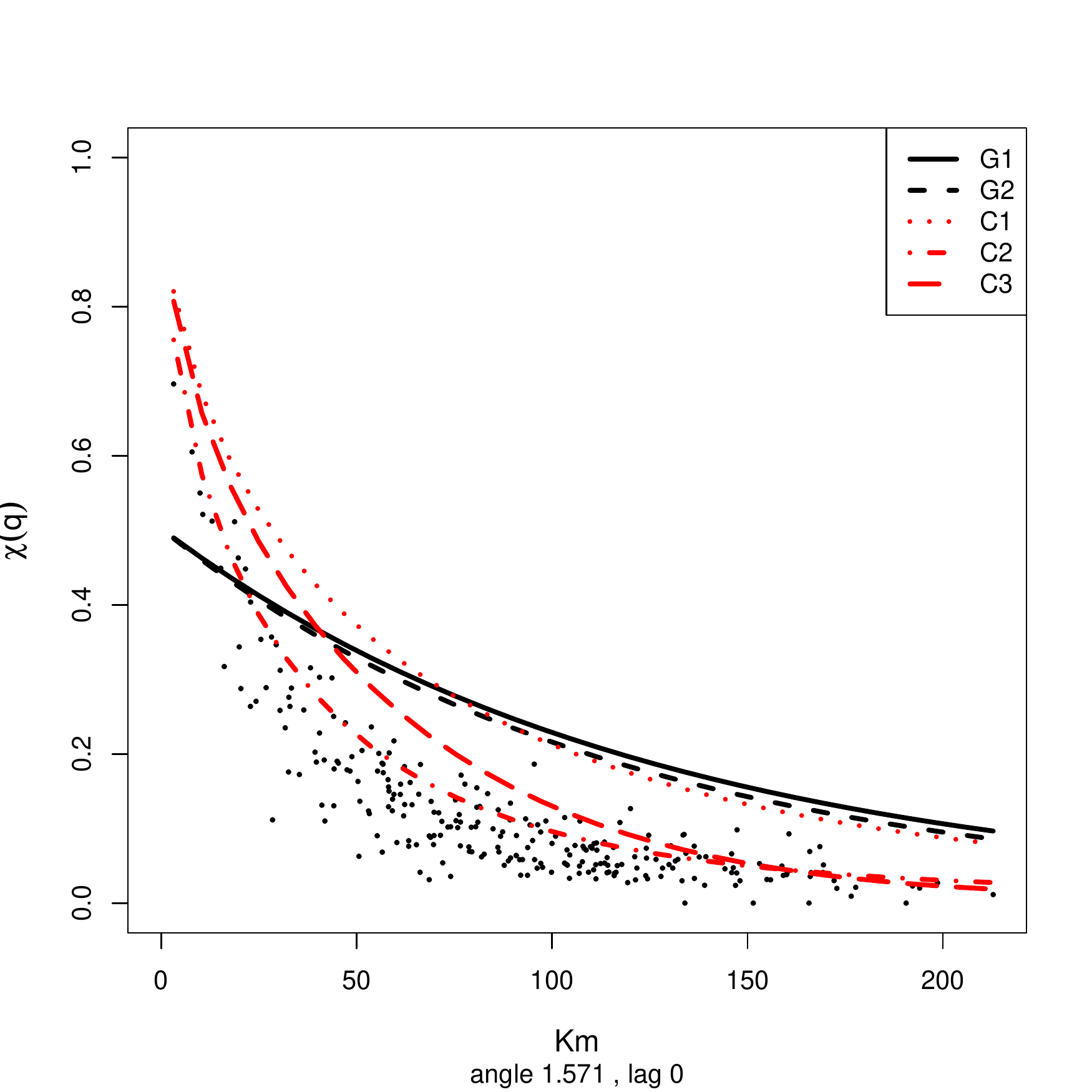}
		\\
		\includegraphics[width=0.3\linewidth]{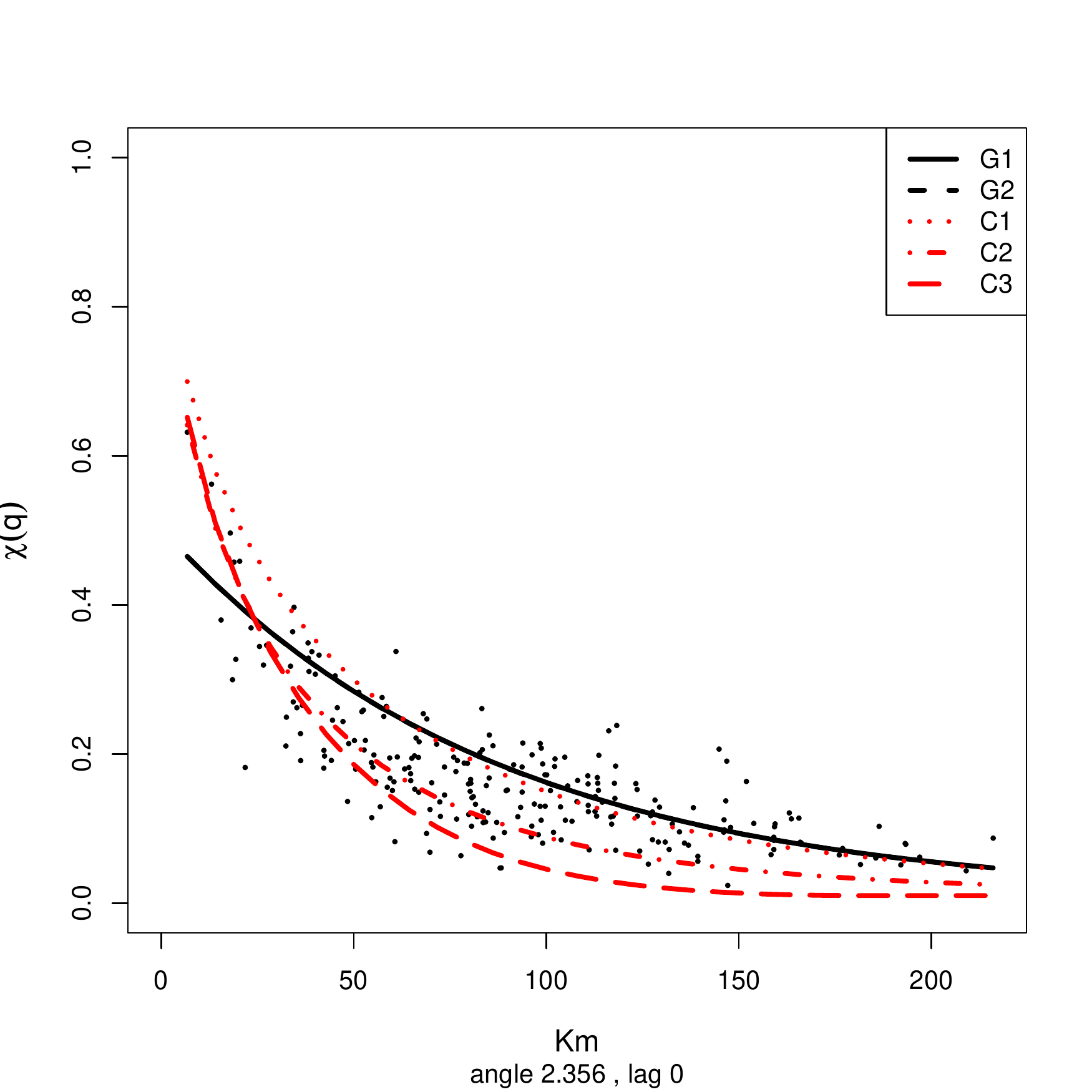}
		&
		\includegraphics[width=0.3\linewidth]{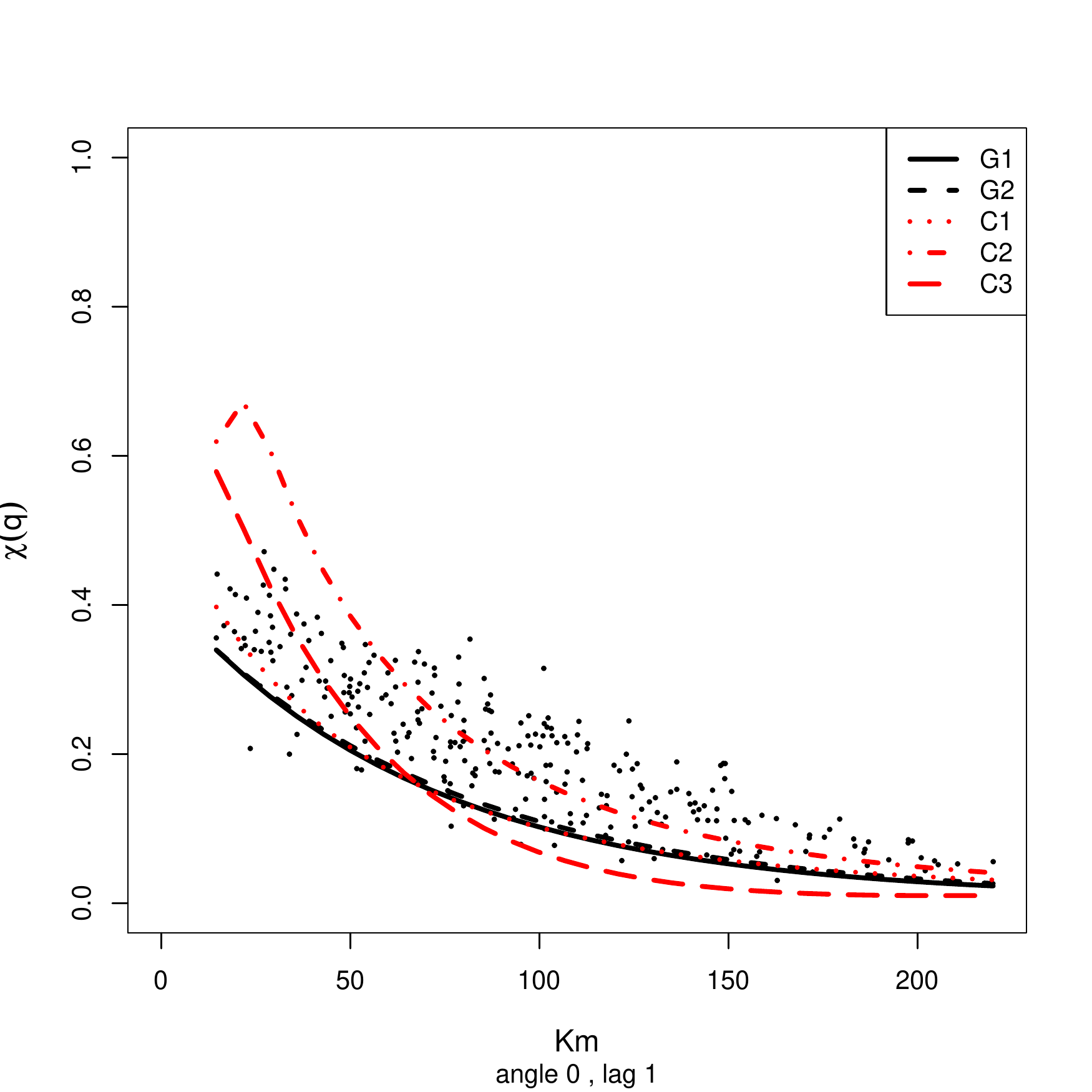}
		&
		\includegraphics[width=0.3\linewidth]{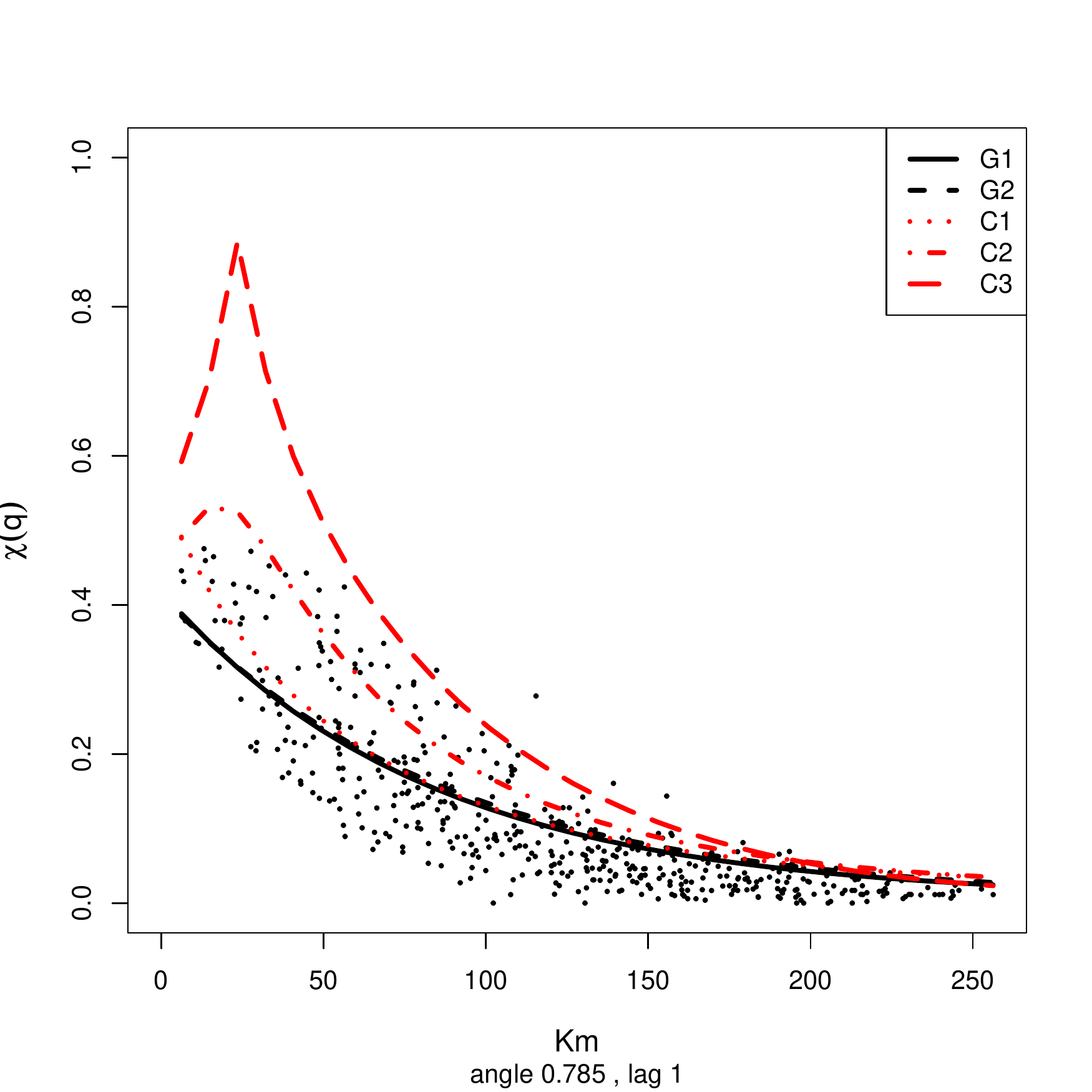}
		\\
		\includegraphics[width=0.3\linewidth]{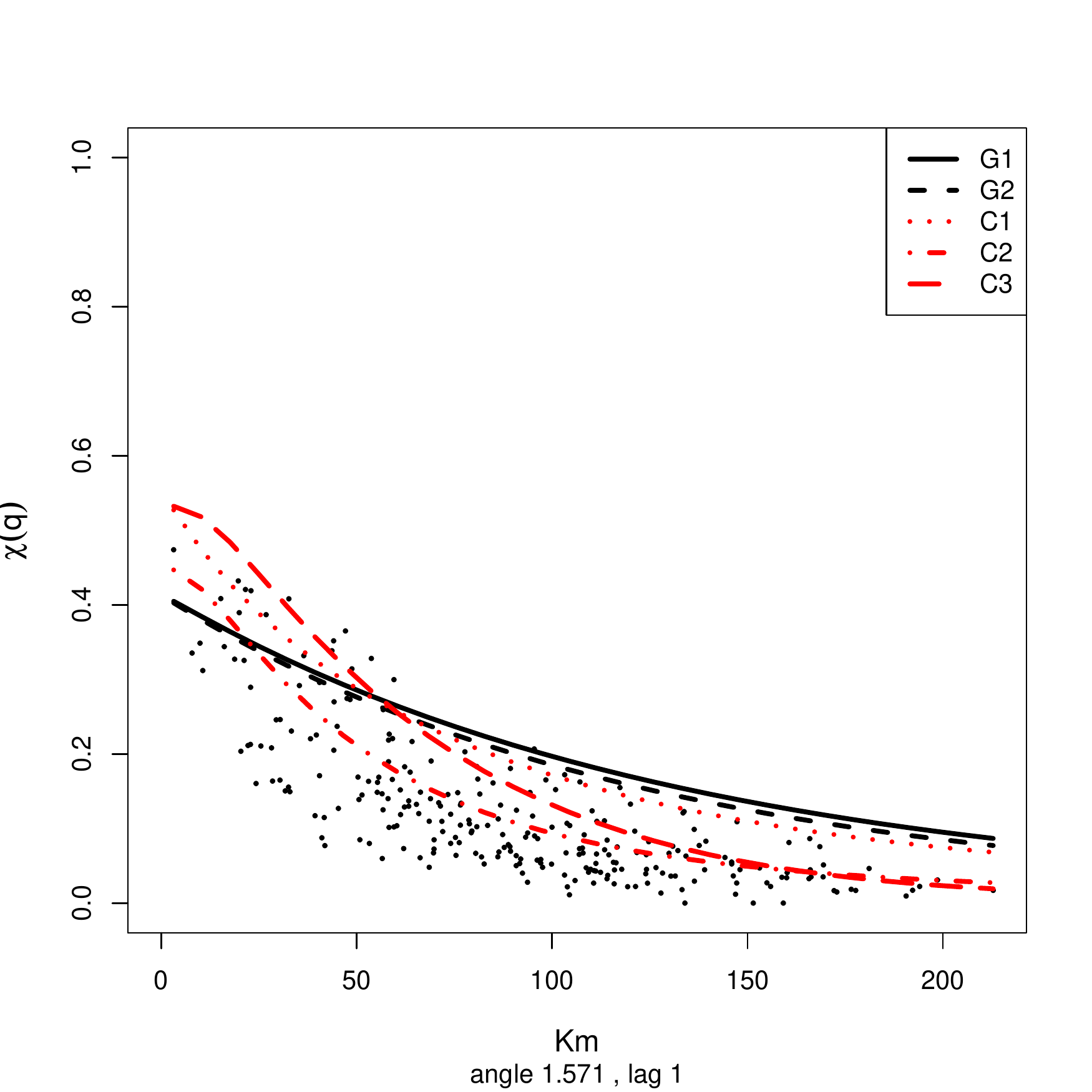}
		&
		\includegraphics[width=0.3\linewidth]{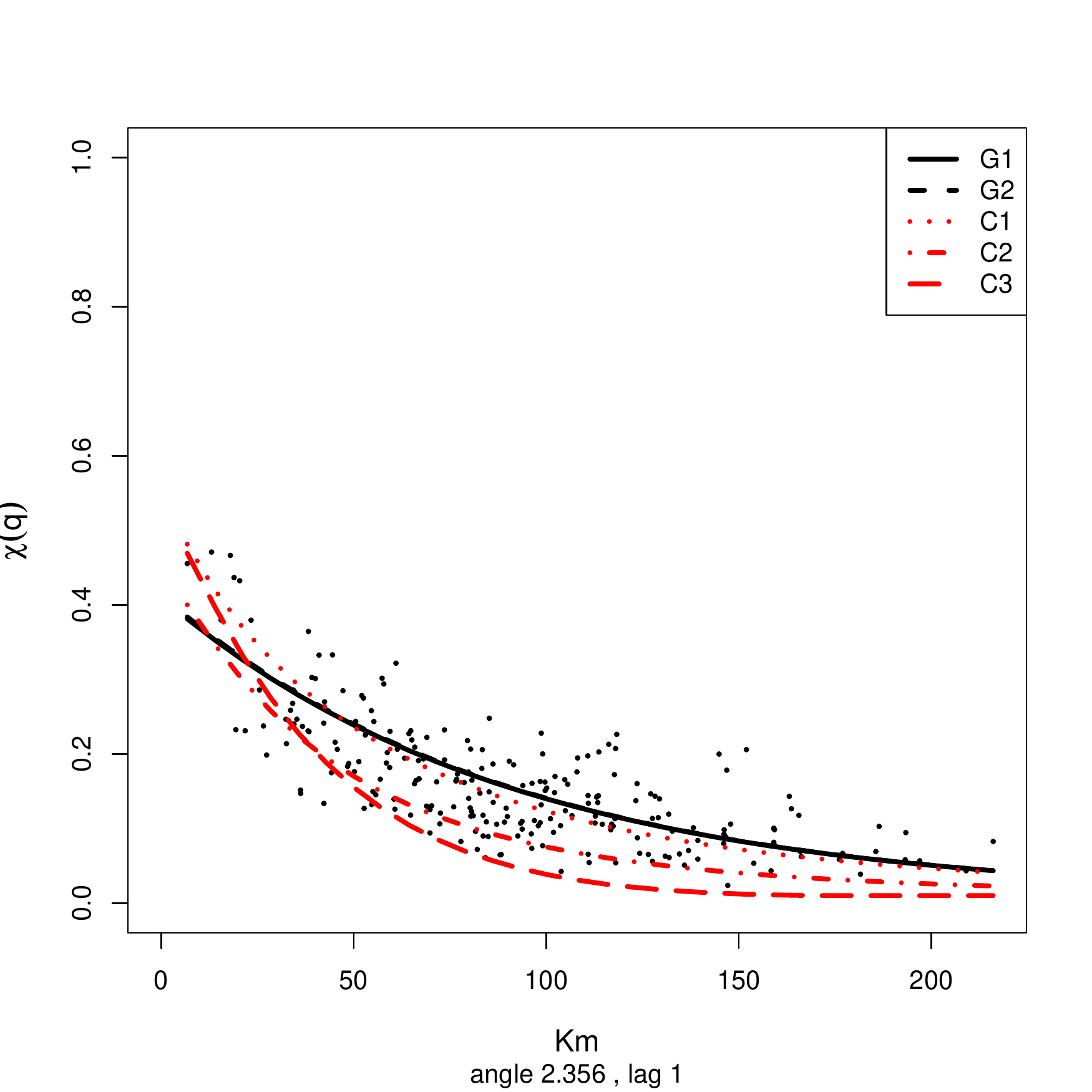}
	\end{tabular}
	
	\caption{Estimated probabilities $\Pr(Z(s,t) > q | Z(s',t') > q)$ along different directions (expressed in radians) and at different temporal lags  for the precipitation data. Dotted points correspond to empirical estimates. The value $q$ is fixed to the empirical  \textit{$99\%$} quantile. }
	\label{fig:gof}
\end{figure}

\end{document}

%% file: authors.tex
\author{Jean-No\"el Bacro \textsuperscript\textdagger
	, Carlo Gaetan \textsuperscript\textdaggerdbl, 
	Thomas Opitz \textsuperscript \textparagraph, 
	Gwladys Toulemonde\textsuperscript{\textdagger \textasteriskcentered} \\
	\textsuperscript\textdagger IMAG, Universit\'e de Montpellier, CNRS, Montpellier, France\\
	\textsuperscript\textdaggerdbl DAIS, Universit\`a  Ca' Foscari di Venezia, Venice, Italy\\
	\textsuperscript\textparagraph BioSP, INRA, Avignon, France\\
	\textsuperscript \textasteriskcentered  INRIA, Project-team LEMON
}

%% file: abstract.tex
\begin{abstract}
\noindent The statistical modeling of space-time extremes in environmental applications is key to understanding complex dependence structures in original event data and to generating realistic scenarios for impact models. In this context of high-dimensional data, we propose a novel hierarchical model for high threshold exceedances  defined over continuous space and time by embedding a space-time Gamma process convolution for the rate of an exponential variable, leading to asymptotic independence in space and time. Its physically motivated  anisotropic dependence structure is based on geometric objects moving through space-time according to a velocity vector.  We demonstrate that inference based on weighted pairwise likelihood is fast and accurate. 
The usefulness of our model is illustrated by an application to  hourly  precipitation data from a study region in Southern France, where it clearly improves on an alternative  censored Gaussian space-time random field model.  While classical limit models based on  threshold-stability fail to appropriately capture relatively fast joint tail decay rates between asymptotic dependence and classical independence,  strong empirical evidence from our application and other recent case studies  motivates the use of more realistic asymptotic independence models such as ours. 
\end{abstract}

%% file: introduction.tex
\section{Introduction}
 The French Mediterranean area is subject to heavy rainfall events occurring mainly in the fall season.  Intense precipitation often leads to flash floods, which can be defined as a sudden strong rise of the water level. Flash floods often cause fatalities and important material damage. In the literature, such intense rainfalls are often called flood-risk rainfall 
\citep{Carreau:Bouvier:2016}; characterizing their spatio-temporal dependencies is key to understanding these processes. In this paper, we consider a large data set of hourly precipitation measurements from a study region in Southern France.  We tackle the challenge of proposing a  physically interpretable statistical space-time model for high threshold exceedances, which aims to capture the complex dependence and time dynamics of the data process. 

Fueled by important environmental applications during the last decade, the statistical modeling of spatial extremes has undergone a fast evolution. A  shift from maxima-based modeling to approaches using threshold exceedances can be observed over recent years, whose reasons lie in the capability of thresholding techniques to exploit more information from the data and to explicitly model the original event data.  A first overview of approaches to modeling maxima is due to \cite{Davison:Padoan:Ribatet:2012}. A number of hierarchical models  based on latent Gaussian processes   \citep{Casson:Coles:1999,Gaetan:Grigoletto:2007,Cooley:Nychka:Naveau:2007,Sang:Gelfand:2009} have been proposed, but they  may be criticized for relying on the rather rigid Gaussian dependence with very weak dependence in the tail, while the lack of closed-form marginal distributions makes interpretation difficult and  frequentist inference cumbersome.  Max-stable random fields
\citep{Smith:1990, Schlather:2002, Kabluchko:Schlather:deHaan:2009, Davison:Gholamrezaee:2012,Reich:Shaby:2012, Opitz:2013} are the natural limit models for maxima data and have spawned a very rich literature,  from which the model of \citet{Reich:Shaby:2012} stands out for its hierarchical construction simplifying high-dimensional multivariate calculations and Bayesian inference. Generalized Pareto processes \citep{Ferreira.deHaan.2014,Opitz.al.2015,Thibaud.Opitz.2015} are the  equivalent limit models for threshold exceedances. However, the asymptotic dependence stability in  these limiting processes for maxima and threshold exceedances  has a tendency to be overly restrictive when asymptotic dependence strength decreases at high levels and may vanish ultimately in the case of asymptotic independence.  The results from the empirical spatio-temporal exploration of our French rainfall data in Section~\ref{sec:explore} are strongly in favor of asymptotic independence, which appears to be characteristic for many  environmental data 
sets
\citep{Davison:Huser:Thibaud:2013, Thibaud:Mutzner:Davison:2013,Tawn.al.2018} and may arise from physical laws such as the conservation of mass. This has motivated the development of more flexible dependence models, such as  max-mixtures of max-stable and asymptotically independent processes 
\citep{Wadsworth:Tawn:2012, Bacro:Gaetan:Toulemonde:2016} or max-infinitely divisible constructions \citep{Huser.al.2018} for maxima data, or Gaussian scale mixture processes \citep{Opitz.2016,Huser.al.2017} for threshold exceedances,  capable to  accommodate asymptotic dependence, asymptotic independence and Gaussian dependence with a smooth transition. Other flexible spatial constructions involve marginally transformed Gaussian processes \citep{Huser.Wadsworth.2018}. Such threshold models can be viewed as part of the wider class of copula models \citep[see][for other examples]{Bortot:Coles:Tawn:2000,Davison:Huser:Thibaud:2013} typically combining univariate limit distributions with dependence structures that should ideally be  flexible and relatively easy to handle in practice. 

Statistical inference is then often carried out assuming   temporal independence in  measurements typically observed at spatial sites at regularly spaced time 
intervals. However, developing  flexible space-time modeling for extremes is crucial for characterizing the temporal persistence of  extreme events spanning several time steps; such models are  important for short-term prediction in applications such as the forecasting of  wind power and atmospheric pollution,   and for extreme event scenario generators providing inputs to impact models, for instance in  hydrology and agriculture.  Early spatio-temporal models for rainfall were proposed in the 1980s \citep[and the references therein]{Rodriguez:Cox:Isham:1987, Cox:Isham:1988} and exploit the idea that storm events give rise to a cluster of rain cells, which are represented as cylinders in space-time. 
Currently, only few statistical space-time models for extremes are available. \cite{Davis:Mikosch:2008} consider extremal properties of  heavy-tailed moving average processes where coefficients and the white-noise process depend on space and 
time, but their work was not focused on practical modeling.   \citet{Sang:Gelfand:2009} propose a hierarchical procedure for maxima data but limited to latent Gauss--Markov random fields.  
\citet{Davis:Kluppelberg:Steinkohl:2013,Davis:Kluppelberg:Steinkohl:2013b}  extend the  widely used class of Brown--Resnick  max-stable processes to the space-time framework and propose pairwise likelihood inference.  Spatial max-stable processes with random set elements have been proposed by \citet{Schlather:2002, 
Davison:Gholamrezaee:2012}, and \cite{Huser:Davison:2014} have fitted a space-time version to threshold exceedances of 
hourly 
rainfall data through pairwise censored 
likelihood. \cite{Huser:Davison:2014} model storms as discs of random radius moving at a 
random velocity for a random 
duration, leading to randomly centered space-time cylinders; our models developed in the following rely on similar geometric representations. A Bayesian approach based on spatial skew-$t$ random fields  with a random set element and temporal autoregression was proposed by \citet{Morris.al.2017}. 
The aforementioned space-time models may capture asymptotic 
dependence or exact  independence at small distances but are unsuitable for dealing with residual dependence in  asymptotic independence. 
In this paper,  we propose a novel approach to space-time modeling of asymptotically 
independent data to avoid the tendency of max-stable-like models to potentially strongly overestimate joint extreme risks.  In a similar context, \cite{Nieto:Huerta:2017} have recently proposed a spatio-temporal Pareto model for heavy-tailed  
data on spatial lattices, generalizing the temporal  latent process model of  \cite{Bortot:Gaetan:2014} to space-time. 

Our model provides a hierarchical formulation for modeling spatio-temporal exceedances over 
high thresholds. It is defined over a continuous space-time domain and allows for a 
physical interpretation of 
extreme events spreading over space and time. Strong motivation also comes from \cite{Bortot:Gaetan:2014} by developing a generalization of their latent temporal process. Alternatively, our latent process could be viewed as a space-time 
version of the temporal trawl processes introduced by \cite{Barndorf:Lunde:Shepard:Veraat:2014} and exploited for extreme values  by \cite{Noven:Veraart:Gandy:2015}, with spatial extensions recently proposed by \citet{Opitz.2017}.
 Our approach is based on  representing a  generalized Pareto distribution as a Gamma mixture of an 
exponential 
distribution, enabling us to keep easily tractable marginal distributions which remain coherent  with univariate extreme 
value theory. We use a kernel convolution of a  space-time Gamma random process \citep{Wolpert:Ickstadt:1998} based on influence zones defined 
as cylinders with an ellipsoidal basis to generate anisotropic spatio-temporal dependence in exceedances.  We then develop statistical inference based on  pairwise likelihood.

The paper is structured as follows. Our hierarchical 
model with a detailed description of its two stages and marginal transformations is developed  in Section~\ref{sec:hierarchical-model}.  Space-time Gamma random fields  are presented in Section~\ref{sec:space-time} where we also propose the construction and formulas for the space-time objects used for kernel convolution.  Section \ref{sec:dependence} characterizes tail dependence behavior in our new model yielding asymptotic independence in  space and time. Statistical inference of model parameters is addressed in Section \ref{sec:composite-likelihood} based on a pairwise log-likelihood for the observed censored excesses. We show good estimation performance  through a simulation study presented in Section~\ref{sec:simulation} involving two scenarios of different  complexity. 
In Section \ref{sec:application}, we focus on the  dataset and explore in detail how our fitted space-time model captures spatio-temporal extremal dependence in hourly precipitation.
Since a natural choice of a  reference model  for asymptotically 
independent data is to use threshold-censored space-time Gaussian processes,  we show the good relative performance of  our model by comparing it to such alternatives. A discussion of our modeling approach with some potential future extensions closes the paper in Section \ref{sec:conclusions}.

%% file: hierarchical-model.tex
\section{A hierarchical  model for  spatio-temporal exceedances}
\label{sec:hierarchical-model}
When dealing with exceedances of a random variable $X$ above a high threshold $u$, univariate extreme value theory suggests using the limit distribution of  Generalized Pareto (GP) type. The GP cumulative distribution function (cdf) is defined for any $y>0$ by
\begin{equation}
\label{gpd}
GP(y;\sigma,\xi)=
1-\left(1+\xi \frac{y}{\sigma}\right)_+^{-(1/\xi)},
\end{equation}
 where  $(a)_+=\max(0,a)$, $\xi$  is a shape 
parameter  and $\sigma$  a positive scale parameter. The sign of $\xi$  characterizes the maximum domain of attraction of the distribution of $X$: $\xi>0$ corresponds to the Fr\'echet domain of attraction  while $\xi=0$ and $\xi<0$ correspond  to the Gumbel and Weibull ones, respectively. 

When $\xi>0$, the GP distribution can be expressed as a Gamma mixture for the rate of the exponential distribution  \citep[p.157]{ Reiss:Thomas:2007}, i.e. 
\begin{equation}\label{eq:gpmixing}
V| \Lambda\sim {\mbox{Exp}}(\Lambda), \quad \Lambda\sim{\mbox{Gamma}}(1/\xi,\sigma/\xi) \quad\Rightarrow\quad V\sim GP(\,\cdot\,;\sigma,\xi),
\end{equation}
where $\mathrm{Exp}(b)$ refers to the Exponential distribution with rate $b>0$  and $\mathrm{Gamma}(a,b)$ to the Gamma distribution with shape $a>0$  and rate $b>0$. 
Based on this hierarchical structure, we will here  develop a stationary space-time construction for modeling exceedances  over a high threshold, which possesses  marginal GP distributions for the strictly positive excess above the threshold. 

\subsection{First stage: generic hierarchical space-time structure}  
We consider a stationary space-time  random field  $Z=\{Z(x),x \in \mathcal{X}\}$   with
$x=(s,t)$ and $\mathcal{X}=\mathbb{R}^2\times \mathbb{R}^+$, such that $s$ indicates spatial location and $t$ time. Without loss of generality, we assume that the margins $Z(x)$ belong to the Fr\'echet domain of attraction with positive shape parameter $\xi$. To infer the tail behavior of $\{Z(x)\}$, we focus on values exceeding a fixed  high threshold $u$, and we consider the exceedances over $u$, 
\begin{equation}
\label{eq:def-exceedance}
Y(x)= (Z(x)-u)\cdot  \mathbf{1}_{(u,\infty)}(Z(x)).
\end{equation}
Standard results from extreme value theory \citep{de_Haan:Ferreira:2006} establish the $GP$ distribution  with $\xi>0$ in \eqref{gpd} as the limit of suitably renormalized positive threshold exceedances in \eqref{eq:def-exceedance}, such that it represents a natural model for the values $Y(x)>0$.
Following \cite{Bortot:Gaetan:2014}, we use the representation of the $GP$ distribution as a Gamma mixture  of an exponential distribution  to formulate  a two-stage  model  that induces spatio-temporal dependence arising  in both the exceedance indicators $\mathbf{1}_{(u,\infty)}(Z(x))$ and   the positive excesses $Z(x)-u>0$  by integrating space-time dependence in a latent Gamma component. A key feature of our model is that it naturally links exceedance probability to the size of the excess and therefore provides a joint space-time structure of the zero part and the positive part in the zero-inflated distribution of $Y(x)$.
 
In the first stage, we condition on a latent space-time random field $\{\Lambda(x)\}$ with marginal distributions $\Lambda(x)\sim\textrm{Gamma}(\alpha,\beta)$ and assume that 
%
%
\begin{subequations}
\label{eq:model}	
	\begin{align}
	Y(x)\mid \left[\Lambda(x), Y(x)>0\right]&\sim {\mbox{Exp}}\left(\Lambda(x)\right), \\
	\Pr(Y(x)>0\mid \Lambda(x))&= e^{-\kappa \Lambda(x)},
	\end{align}
\end{subequations}
where  $\kappa >0$ is a parameter controlling the rate of 
upcrossings of the threshold. The resulting marginal distribution of  $Y(x)$ conditionally on $Z(x) > u$  corresponds to  the GP distribution,  and the unconditional marginal cdf 
of $Y(x)$ is 
\begin{equation}
\label{eq:censored-distribution}
F(y;\sigma,\xi)=
\left\{
\begin{array}{lc}
p & \textrm{for } y=0,\\
p+(1-p)GP(y;\xi,\sigma) & \textrm{for } y >0,
\end{array}
\right.
\end{equation}
with shape parameter 
 $\xi=1/\alpha$, scale parameter  $\sigma= (\kappa+\beta)/\alpha$, and with $1-p$ the probability of an exceedance over $u$, i.e. $\Pr(Z(x)> u)=\Pr(Y(x)>0)=1-p$. The probability  of exceeding $u$,   
 \begin{equation}\label{eq:marginal}
\Pr(Z(x)>u)=\E\left(\Pr(Y(x)>0|\Lambda(x))\right) =\E\left(e^{-\kappa\Lambda(x)}\right)=
 \left(\frac{\beta}{\kappa+\beta}\right)^{\alpha}
\end{equation}
depends on $\kappa$ and corresponds to the Laplace transform of $\Lambda(x)$ evaluated at $\kappa$.
The constraint $\xi >0$ is not restrictive for dealing with precipitation in the French Mediterranean area, which is known to be heavy-tailed. For general modeling purposes,  we can relax this assumption by  following  \citet{Bortot:Gaetan:2016}: we consider a marginal transformation within the class of GP distributions for threshold exceedances, for which we suppose that  $\alpha=1$  and  $\beta=1$ for identifiability. By transforming $Y(x)$  through the probability integral transform
\begin{equation}
\label{eq:transf}
g(y)=GP^{-1}(GP(y;1,1+\kappa);\sigma^*,\xi^*) = 
(\sigma^*/{\xi^*})\left\{\left(1+\frac{y}{\kappa+1}\right)^{\xi^*}-1\right\} \end{equation}
with parameters $\xi^* \in  \mathbb{R}$ and $\sigma^* >0$ to be estimated, we get a marginally transformed random field $
Y^*(x)= g(Y(x))$
which satisfies   
$Y^*(x)\sim GP(\,\cdot\,;\xi^*,\sigma^*)$, conditionally on $Y^*(x)>0$. Notice that it is straightforward to develop extensions with nonstationary marginal excess distributions by injecting response surfaces $\sigma^\star(x)$ and $\xi^\star(x)$ into \eqref{eq:transf}.
Moreover, nonstationarity could be introduced into the latent Gamma model  \eqref{eq:model}  in different ways. If $\kappa=\kappa(x)$ depends on $x$ or other covariate information, exceedance probabilites become nonstationary. If Gamma parameters $\alpha=\alpha(x)$ and $\beta=\beta(x)$ depend on covariates, then the GP margins in $Y(x)$ become nonstationary. Finally, one could combine the two previous nonstationary extensions.

%% file: space-time.tex
\label{sec:space-time}
\subsection{Second stage: space-time dependence with Gamma random fields}
Spatio-temporal dependence is introduced by means of a space-time stationary random field $\{\Lambda(x), x\in \mathcal{X} \}$ with  $\mathrm{Gamma}(\alpha,\beta)$ marginal distributions. In principle, we could use an arbitrarily wide range of models with any kind of space-time dependence structure, for instance by  marginally transforming a space-time Gaussian random field using the copula idea \citep{Joe:1997}.  However, we here aim to propose a construction where Gamma marginal distributions arise naturally without applying rather artificial marginal transformations. 
Inspired by the Gamma process convolutions of \citet{Wolpert:Ickstadt:1998}, we develop a space-time Gamma convolution process with Gamma marginal distributions.  The kernel shape in our construction allows for a straightforward interpretation of the dependence structure, and it  offers  a physical interpretation of real   phenomena such as mass and particle transport. Moreover, we obtain simple analytical formulas for the bivariate distributions, which facilitates statistical inference, interpretation and the characterization of joint tail properties. 

We fix $\mathcal X=\mathbb{R}^{3}$ and consider $A\in  \mathcal{B}_b(\mathcal X)$, a subset of $\mathcal X$
belonging to the $\sigma-$field $\mathcal{B}_b(\mathcal X)$ restricted to bounded sets
of $\mathcal X$. A Gamma random field  $\Gamma(dx)$ \citep{Ferguson;1973} is a non negative random measure  defined on $\mathcal X$  characterized by a base measure $\alpha(dx)$ and a rate parameter $\beta$ such that
\begin{enumerate}
	\item $\Gamma(A):= \int_A \Gamma(dx) \sim \textrm{Gamma}(\alpha(A),\beta)$, with $\alpha(A) :=\int_A \alpha(dx)$;
	\item for any $A_1,A_2\in  \mathcal{B}_b(\mathcal X)$ such that	$A_1\cap A_2=\emptyset$, $\Gamma(A_1)$ and $\Gamma(A_2)$ are independent random variables. 
\end{enumerate}
The calculation of important formulas in this paper requires the Laplace exponent of the random measure given as
$$
\mathcal{L}(\phi) := -\log\E\left(\exp\left\{-\int \phi(x)\Gamma(dx)\right\}\right)
= \int_{\mathcal X} \log\left\{1+\frac{\phi(x)}{\beta}\right\}\alpha(dx)
$$
where $\phi$ is any positive measurable function; in our case, it will represent the kernel function (see the Appendix section~\ref{sec:Laplace}). 
We propose to model $\{\Lambda(x), x\in \mathcal{X} \}$ as a convolution using a 3D indicator kernel $K(x, x')$ with an indicator set of finite volume used to convolve the Gamma random field $\Gamma(dx)$ \citep{Wolpert:Ickstadt:1998}, i.e.,  $\Lambda(x)=\int K(x,x')\Gamma(dx')$.
The shape of the kernel can be very general (although non indicator kernels usually do not lead to Gamma marginal distributions), and particular choices may lead to nonstationary random fields, or to stationary random fields with given dependence properties such as  full symmetry, separability or independence beyond  some spatial distance or temporal lag.
In order to limit model complexity and computational burden to a reasonable amount,  we use  the indicator kernel
$K(x,x')=\mathbf{1}_A(x-x')$, for $A\in \mathcal{B}_b(\mathcal X)$,  
where $A$ is given as a slated elliptical cylinder, defining  a $3$-dimensional set $A_x$ that moves through  $\mathcal{X}$ according to some velocity vector. More precisely, let $E(s,\gamma_1,\gamma_2,\phi)$ be  an ellipse 
centered at $s=(s_1,s_2)\in \mathbb{R}^2$ (see  Figure \ref{fig:examples}-(a)), whose axes are rotated counterclockwise by the angle $\phi$ with respect  
to the coordinate axes, whose semi-axes' lengths in the rotated coordinate system are $\gamma_1$ and $\gamma_2$, respectively. 
A physical  interpretation is that the ellipse describes the spatial influence zone of a storm centered at $s$. For the temporal dynamics, we assume that the ellipses (storms) $E(s,\gamma_1,\gamma_2,\phi)$ move through space with a velocity  $\omega=(\omega_1,\omega_2)
\in \mathbb{R}^2$ for a duration $\delta>0$. The volume of the intersection of two slated elliptical cylinders (see Figure \ref{fig:examples}-(b)) is  given by
$$
V(s,t,s',t')=(\delta-|t-t'|)_+ \times \nu_2(E(s,\gamma_1,\gamma_2,\phi)\cap E(\tilde{s},,\gamma_1,\gamma_2,\phi))
$$
where $\tilde{s}=(\tilde{s_1},\tilde{s_2})$ with
$\tilde{s_i}=s_i'-|t'-t|\times \omega_i$, $i=1,2$, where $\nu_d(\cdot)$
is the Lebesgue measure on $\mathbb{R}^d$.

For two fixed locations, the strength of dependence in the random field $\Lambda(x)$ is an increasing monotone function of the intersection volume;  other choices of $A$ are possible,  provided that we are able to calculate efficiently the volume of the intersection. 
To efficiently calculate the ellipse intersection area, we use an approach for finding the overlap area between two ellipses, which does not rely on proxy curves; see \citet{Hughes:Chraibi:2012}\footnote{The code  is open source and can be downloaded from  \texttt{http://github.com/chraibi/EEOver}.}.

In the sequel, we consider the  measure 
\begin{equation}\label{eq:gammanoise}
\alpha(B)=\alpha \nu_d(B)/\nu_d(A), \quad B\in  \mathcal{B}_b(\mathcal X).
\end{equation}
It follows that
$\Lambda(x)\sim \textrm{Gamma}(\alpha,\beta)$, as required for model (\ref{eq:model}). Exploiting the formulas of the Appendix 
section~\ref{sec:Laplace}, the univariate Laplace transform of $\Lambda(x)$  is
\begin{equation}
\label{lp1}
LP_x^{(1)}(v):= \E\left(e^{-v \Lambda(x)}\right)=\left(\frac{\beta}{v+\beta}\right)^{\alpha},
\end{equation}
and  the  bivariate Laplace transform of 
$\Lambda(x)$ and $\Lambda({x'})$ is
\begin{eqnarray}\label{lp2}
LP_{x,x'}^{(2)}(v_1,v_2)&:= &\E\left(e^{-v_1 \Lambda(x)-v_2\Lambda({x'})}\right)\nonumber\\
&=&
\left(\frac{\beta}{v_1+\beta}\right)^{\alpha(A_{x}\backslash A_{x'})}
\left(\frac{\beta}{v_1+v_2+\beta}\right)^{\alpha(A_{x}\cap A_{x'})}
\left(\frac{\beta}{v_2+\beta}\right)^{\alpha(A_{x'}\backslash A_{x})}.
\end{eqnarray}

This model for $\Lambda(x)$ is stationary, but nonstationarity in Gamma marginal distributions and/or dependence can be generated by using nonstationary indicator sets $A_{x}$ whose size and shape depends on $x$. More general sets $A_x$ with finite Lebesgue volume $\nu_3(A_x)$ could be used for constructing $\Lambda(x)=\Gamma(A_x)$. In all cases, the intersecting volume $\nu_3(A_{x_1}\cap A_{x_2})$ tends to zero  if  $\|x_2-x_1\|\rightarrow\infty$, which establishes the property of $\alpha$-mixing over space and time for the processes $\Lambda(x)$ and $Y(x)$. This property is paramount to ensure consistency and asymptotic normality in the pairwise likelihood estimation that we consider in the following (see \cite{Huser:Davison:2014}).


%% file: dependence.tex
\section{Joint tail behavior of the hierarchical process}
\label{sec:dependence}
Extremal dependence in a bivariate random vector  $(Z_1,Z_2)$  can be explored based on 
the tail behavior of the conditional distribution $\Pr(Z_1>F_1^{\leftarrow}(q)| Z_2>F_2^{\leftarrow}(q))$ as $q$ tends to $1$,  where 
$F_i^{\leftarrow}$, $i=1,2$ denotes the generalized inverse distribution functions of $Z_i$ \citep{Sibuya:1960, Coles:Heffernan:Tawn:1999}. 
The random vector  $(Z_1,Z_2)$  is said to be asymptotically dependent if a positive limit $\chi$, referred to as the tail correlation coefficient, arises:
$$
\chi(q):= 
\frac{\Pr(Z_1>F_1^{\leftarrow}(q),Z_2>F_2^{\leftarrow}(q))}{ \Pr( Z_2>F_2^{\leftarrow}(q))}\rightarrow\chi > 0, \qquad q \rightarrow 1^{-}.
$$
The case  $\chi=0$ characterizes asymptotic independence. 

To obtain a finer characterization of the joint tail decay rate under asymptotic independence, faster than the marginal tail decay rate,  \cite{Coles:Heffernan:Tawn:1999} have introduced the $\overline \chi$ index defined  through the limit relation
$$
\bar{\chi}(q):=
\frac{2\log \Pr (Z_2>F_2^{\leftarrow}(q))}{\log {\Pr}(Z_1>F_1^{\leftarrow}(q),Z_2>F_2^{\leftarrow}(q))}-1
\rightarrow\bar{\chi}\in (-1,1], \qquad  q\rightarrow 1^-.
$$
Larger values of $|\bar{\chi}|$ correspond to stronger dependence. We now show that $\{Z(x), x \in \mathcal{X} \}$ is an asymptotic independent process, i.e., for all pairs $
(x,x')\in \mathcal{X}^2$ with $x\not=x'$ the bivariate random vectors $(Z(x),Z(x'))$ are asymptotically independent.

Owing to the stationarity of the process, 
it is easy to show that for any $(x,x')\in 
\mathcal{X}^2$, $x\not = x'$ and for values $v$ exceeding a threshold $u \geq 0$, we get 
\begin{eqnarray*}
\Pr(Z(x)>v)&=&LP_x^{(1)}(v-u+\kappa)
\\
&=&\left(1+\frac{v-u+\kappa}{\beta}\right)^{-\alpha(A_x)}
\end{eqnarray*}
and 
\begin{eqnarray*}
\Pr(Z(x)>v,Z(x')>v)&=&LP_{x,x'}^{(2)}(v-u+\kappa,v-u+\kappa)\\
&=&\left(1+\frac{v-u+\kappa}{\beta}\right)^{-\alpha(A_{x}\backslash 
A_{x'})}\left(1+\frac{2v-2u+2\kappa}{\beta}\right)^{-\alpha(A_{x}\cap A_{x'})}
\\
&&\times \left(1+\frac{v-u+\kappa}{\beta}\right)^{-\alpha( A_{x'}\backslash A_{x})}.
\end{eqnarray*}
To simplify notations, we set  $c_0:=\alpha(A_x)$, $c_1:=\alpha(A_{x}\backslash 
A_{x'})$, $c_2:=\alpha(A_{x}\cap A_{x'})$ , $c_3:=\alpha( A_{x'}\backslash A_{x})$, such that 
$c_1=c_3=c_0-c_2\geq 0$ and $c_1+2c_2+c_3=2c_0$. For $c_2=0$ characterizing disjoint indicator sets $A_x$ and $A_{x'}$, it is clear that  $Z(x)$ and $Z(x')$ are independent.
Now, assume $u=0$ without loss of generality and $x\not = x'$; then,
\begin{eqnarray*}
\chi_{x,x'}(v)&:=&\frac{\Pr(Z(x)>v,Z(x')>v)}{\Pr(Z(x')>v)}
\\
 &=&\left( 1+\frac{2v+2\kappa}{\beta}\right)^{-c_2} \left( 1+\frac{v+\kappa}{\beta}\right)^{-c_1-c_3+c_0}\\
&=& \left( 1+\frac{2v+2\kappa}{\beta}\right)^{-c_2}\left( 1+\frac{v+\kappa}{\beta}\right)^{2c_2-c_0}\\
&\sim& {2}^{-c_2}\left(\frac{v}
{\beta}\right)^{c_2-c_0},\qquad \mbox{for large $v$}.
\end{eqnarray*}
Since $c_2<c_0$, we obtain 
$$\chi_{x,x'} =0.$$
We conclude that $Z$ is 
an asymptotic independent process. 

To characterize the faster joint tail decay, we calculate
\begin{eqnarray*}
\bar{\chi}_{xx'}(v)&:=& 
\frac{2\log {\rm Pr}(Z(x)>v)}{\log {\rm Pr}(Z(x)>v,Z(x')>v)}-1\\
&=&
\cfrac{-2c_0\log \left(1+(v+\kappa)/{\beta}\right)}{-c_1\log  (1+(v+k)/{\beta}) -c_2\log\left( 1+ 2(v+k)/
{\beta}\right)-c_3\log (1+(v+k)/{\beta})}-1\\
&=
&\cfrac{2c_0}{c_1+c_2\cfrac{\log\left( 1+ 2(v+k)/{\beta}\right)}{ \log  \left(1+(v+k)/\beta \right)}+c_3}-1.\\
\end{eqnarray*}
Taking the limit for $v\rightarrow\infty$ yields
$$
\bar{\chi}_{x,x'}=  \cfrac{2c_0}{c_1+c_2+c_3}-1=\cfrac{c_2}{2c_0-c_2},
$$
 which describes the ratio between the  intersecting volume of $A_x$ and $A_{x'}$ and  the volume of the union of these two sets. 
The value of $\bar{\chi}$ confirms the asymptotic independence of the process $Z$. 
A larger intersecting volume between $A_x$ and $A_{x'}$ corresponds to stronger dependence.

%% file: composite-likelihood.tex
\section{Composite likelihood inference}
\label{sec:composite-likelihood}
To infer the tail behavior of the observed data process $\{Z(x)\}$, without loss of generality assumed to have generalized Pareto marginal distributions with shape parameter $\alpha$, we focus on values exceeding a fixed high threshold $u$. 
We let $\theta$ denote the vector of unknown parameters. For simplicity, we assume  that we have  observed the excess values $Y(s_i,t)$ for a factorial design of $S$ locations $s_i$, $i=1,\ldots,S$ and  $T$ times $t=1,\ldots, T$. 

To exploit the tractability of intersecting volumes of two kernel sets, we focus on pairwise likelihood for efficient inference in our high-dimensional space-time set-up. 
The pairwise  (weighted) log-likelihood  adds up the contributions $f (Y(s_i,t),Y(s_j,t+k);\theta)$ of  the censored  observations $Y(s_i,t),Y(s_j,t+k)$ and can be written
\begin{eqnarray}
\mathrm{pl}(\theta)
=\sum_{t=1}^{T} \mathrm{pl}_t(\theta) =\sum_{t=1}^T \sum_{k=0}^{\Delta_T} \sum_{i= 1}^{S}\sum_{j=1}^S\{1-{1}_{\{i \ge j,\, k=0\}} \}\log f (Y(s_i,t),Y(s_j,t+k);\theta) {w_{s_i,s_j}}\label{pln}
\end{eqnarray}
where $w_{s_i,s_j}$ is a weight defined on $[0,\infty)$ \citep{Bevilacqua:Gaetan:Mateu:Porcu:2012, Davis:Kluppelberg:Steinkohl:2013b, Huser:Davison:2014}. We opt for a cut-off weight with $w_{s_i,s_j}= 1$ if
$||s_i-s_j|| \le \Delta_S$  and $0$ otherwise, which bypasses an explosion of the number of likelihood terms and shifts focus to relatively short-range distances where dependence matters most. This also avoids that the pairwise likelihood value (and therefore parameter estimation) is dominated by a large number of intermediate-range distances where dependence has already decayed to (almost) nil.

The contributions $f(Y(x),Y(x');\theta)$ are given by
$$
f(y_1,y_2;\theta)=
\left\lbrace
\begin{array}{ll}
\frac{\partial^2}{\partial v_1\partial v_2}LP^{(2)}_{x,x'}(v_1,v_2) J(y_1) J(y_2) & y_1> 0,y_2 >0
\\
\left( -\frac{\partial}{\partial v_1}LP^{(1)}(v_1)+
\frac{\partial}{\partial v_1}LP_{x,x'}^{(2)}(v_1,v_2)\right) J(y_1)& y_1> 0,y_2 =0
\\
\left( -\frac{\partial}{\partial v_2}LP^{(1)}(v_2)+\frac{\partial}{\partial v_2}LP_{x,x'}^{(2)}(v_1,v_2)\right) J(y_2)
& y_1 =0, y_2 >0 
\\
1-2 LP^{(1)}(v_1)+LP_{x,x'}^{(2)}(v_1,v_2)
& y_1=0,y_2 =0
\end{array}
\right.
$$
with $v_i=(\kappa+1) \left( 1+\xi^* y_i/\sigma^*\right)^{1/ \xi^*}-1$
and $J(y_i)=\frac{\kappa+1}{\sigma^*}\left(1+\frac{\xi^*y_i}{\sigma^*} \right)^{1/ \xi^* -1}$, $i=1,2$. We provide analytical expressions for $LP^{(1)}$ and $LP_{x,x'}^{(2)}$ in the Appendix section~\ref{sec:likelihood}.

Since the space-time random field $\{\Lambda(x)\}$ is temporally $\alpha$-mixing,  the maximum
pairwise likelihood estimator $\widehat{\theta}$ can be shown to be asymptotically normal for large $T$ under mild additional regularity conditions; see Theorem~1 of \citet{Huser:Davison:2014}.
The asymptotic variance is given by the inverse of the Godambe information matrix
$\mathcal{G}(\theta)=\mathcal{H}(\theta)[\mathcal{J}(\theta)]^{-1}\mathcal{H}(\theta)$.
Therefore, standard error evaluation  requires consistent estimation of the matrices $\mathcal{H}(\theta)=\E(- \nabla^2 \mathrm{pl}(\theta))$ and $\mathcal{J}(\theta)=\Var(\nabla \mathrm{pl}(\theta))$. We estimate   $\mathcal{H}(\theta)$ with 
$\widehat{\mathcal{H}}=-\nabla^2 \mathrm{pl}(\widehat {\theta})$
and $\mathcal{J}(\theta)$
through a subsampling technique \citep{Carlstein:1986}, implemented as follows. 
We define $B$ overlapping blocks $D_b\subset\{1,\ldots,T\}$, $b=1,\ldots,B$,  containing   $d_b$ observations; we write $pl_{D_b}$ for the pairwise likelihood (\ref{pln}) evaluated over the block $D_b$. The  estimate of $\mathcal{J}(\theta)$ is
$$
\widehat{\mathcal{J}}=\frac{T}{B}\sum_{b=1}^B \frac{1}{d_b}\nabla \mathrm{pl}_{D_b}(\widehat{\theta})\nabla \mathrm{pl}_{D_b}(\widehat{\theta})^\prime.
$$
The estimates $\widehat{\mathcal{H}}$ and  $\widehat{\mathcal{J}}$ allow us to calculate  the composite likelihood information 
criterion \citep{Varin:Vidoni:2005} 
$$\mathrm{CLIC}=-\mathrm{pl}(\hat \theta)+\mathrm{tr}\{\hat{\mathcal H}^{-1} \hat{\mathcal J} \}
$$
with lower values of $\mathrm{CLIC}$ indicating a better fit.
Similar to \citet{Davison:Gholamrezaee:2012}, we improve the interpretability of \mbox{CLIC} values through rescaling \mbox{CLIC}$^*$=$c$\,\mbox{CLIC} by a positive constant $c$ chosen to give  a pairwise log-likelihood value
 $\mathrm{pl}(\theta)$ comparable to the log-likelihood under independence.

%% file: simulation.tex
\section{Simulation study}
\label{sec:simulation}
We assess the performance of the pairwise composite likelihood estimator through a  small simulation study. For each replication, we consider 
  $S = 30$ randomly chosen sites on $[0,1]\times[0,1]$ observed at time points $t=1,\ldots, T = 2000$. The realizations of the Gamma random field are simulated
  by    adapting  the algorithm of  \citet{Wolpert:Ickstadt:1998b}.
In the simulations, we fix parameters $\xi=1$, $\sigma=10$ and an exceedance probability of $1-p=0.2$. We focus on estimating  dependence parameters while treating the margins as known. For estimation,  we fix the site-dependent threshold $u$  to an empirical quantile of order greater than $p$. Here, we fix $p=0.9$ corresponding to $\kappa=9$.

Two scenarios with different model complexity are considered, involving different specifications of the   cylinder (see Table \ref{tab:sim}).  Scenario A  uses a circle-based cylinder without velocity, while  Scenario B comes with a slated ellipse-based cylinder, yielding  non null velocity.  Technically, the  model in  Scenario A is over-parametrized since the rotation parameter  $\phi$  cannot change the volume of the cylinder.


Model parameters are estimated on $100$ data replications using the composite likelihood approach developed in Section \ref{sec:composite-likelihood}.  We have considered a larger number of replications for some parameter combinations,  but in general  the number of  $100$ replications is enough to satisfactorily illustrate the estimation efficiency. 
The evaluation of $pl(\theta)$ depends on the choice of $\Delta_S$ and $\Delta_T$, where greater values increase the computational cost.
Results in the literature indicate that using as much as computationally possible or all of the pairs will not necessarily lead to an improvement in estimation owing to potential issues with estimation variance (see  \citet{Huser:Davison:2014}, for instance). 
We have considered different values of $\Delta_T$ and have identified
 $\Delta_T=15$ as a good compromise for the estimation quality.  The parameter  $\Delta_S$ has been set to $1$ which is large enough with respect to the spatial domain limits.
Main results  are illustrated in the  boxplots in Figures \ref{fig:boxplotsA} and \ref{fig:boxplotsD}. 

When the cylinder is circle-based, i.e. $\gamma_1=\gamma_2$, and without velocity (scenario A),  the orientation parameter $\phi$  can take any value. In the simulation experiment we estimate all parameters without constraints, such that the optimization algorithm gives also an estimate of $\phi$.
It is reassuring to see in the boxplots of Fig.~2 that  the other parameters are still well estimated.

Results  are fairly good for the scenario B where the velocity is non null. The estimates of the velocity present slightly higher variability,  and the estimation of $\omega_2$ appears  
slightly biased. On the other hand, the duration $\delta$ and  the lengths of the semi-axes of the  ellipse ($\gamma_1$ 
and $\gamma_2$) are still well estimated. {\textit The angle $\phi$ is well defined in scenario B, but it is still estimated with relatively high variability. }
This may seem as disappointing at first glance, but it may be due to the only moderate difference in the length of the semi-axes. To check this conjecture, 
we consider a modified scenario B where the second semi-axis is modified from $\gamma_2=0.3$ to $\gamma_2=0.5$ and other parameters remain unchanged. 
As illustrated by the boxplots in Figure  \ref{fig:phi},  estimation of $\phi$ clearly  improves when the shape of the ellipse departs more strongly from a circular shape.

Even with only a relatively small number of spatial sites and time steps, the simulation study shows that the pairwise composite likelihood 
approach leads to reliable estimates of model parameters that are well identifiable. We underline that results are consistently good whatever the complexity of 
the  scenario. 

%% file: real-data-example.tex
\section{Space-time modeling of hourly precipitation data in southern France}
\label{sec:application}
\subsection{Data}
We apply our hierarchical model  to precipitation extremes observed over a study region
in the South of France. Extreme rainfall events  usually occur  during fall season. They are mainly due to southern winds driving warm and moist air from the Mediterranean sea towards the relatively cold mountainous areas of the Cevennes and  the Alps, leading to a situation which often provokes severe thunderstorms. The data  were provided by M\'et\'eo France  (\texttt{https://publitheque.meteo.fr}). Our dataset is part of a query containing hourly observations at $213$ rainfall stations  for  years 1993 to 2014. To avoid modeling  complex seasonal trends, we keep only data 
 from  the September to  November months,   
 resulting in observations over $54542$ hours.
For model fitting, we consider a subsample of $50$ meteorological stations with  elevations ranging from $2$ to $1418$ meters, for which the observation series contain less than $70\%$ of missing values over the full period. The spatial design of the stations is illustrated in  Figure \ref{fig:map}.
 
 \subsection{Exploratory analysis}
 \label{sec:explore}
 We fit the univariate model (\ref{eq:censored-distribution}) for each station by fixing a threshold $u$ that corresponds to the empirical $99\%$  quantile. We use such a rather high probability value since we have many observations, and there is  a substantial number of zero values such that a high quantile is needed to get into the tail region of the positive values. 
Figure \ref{fig:map} clearly shows that spatial nonstationarity arises in the marginal distributions.
 
 Figure \ref{fig:empchi} displays the results of a bootstrap  procedure in which we calculate estimates of $\chi(q)$ and $\bar{\chi}(q)$ for probabilities $q = 0.99,  0.995$ for pairs $Z(s, t)$, $Z(s, t + h)$ with only temporal lag, and for pairs $Z(s, t)$, $Z(s', t)$ with only spatial lag. The curves for spatial lags are the result of a  smoothing procedure. Confidence bands are based on $200$ bootstrap samples, drawn by the stationary bootstrap \citep{Politis:Romano:1994}. 
Our procedure samples temporal blocks of observations and the block length follows a geometric distribution with  an average  of $20$ days. These plots  support  the assumption of  asymptotic independence at all positive distances and at all positive temporal lags.  Moreover, the strength of tail dependence as measured by the subasymptotic tail correlation value $\chi(q)$ strongly decreases when considering exceedances over increasingly high thresholds, which provides another clear sign of continuously decreasing and ultimately vanishing dependence strength. 
On the other hand, the values of the subasymptotic dependence measure  $\overline{\chi}(q)$ (well adapted to asymptotic independence)  decrease with increasing spatial distances or  temporal lags, but they tend to stabilize at a non zero value. This behavior indicates the presence of residual tail dependence that vanishes only asymptotically. 

\subsection{Modeling spatio-temporal dependence}
While the preceding exploratory analysis has shown that marginal distributions are not stationary, our  model detailed in Section \ref{sec:hierarchical-model} requires a specific type of common marginal distributions. It would indeed be possible to extend the model  to accommodate non stationary patterns (an example can be found in \citet{Bortot:Gaetan:2016})  and to jointly estimate   marginal and dependence parameters. However, our focus here is to illustrate that our modeling strategy is capable to capture complex stationary spatio-temporal dependence patterns at large values, which would render joint estimation of margins and dependence highly intricate. 
Therefore, we fit a GP  distribution separately to each site with thresholds chosen as the empirical $99\%$ quantile. With respect to positive precipitation, this quantile globally corresponds to a probability of  $0.91$, with  a minimum of $0.86$ and maximum of $0.95$ over the $50$  sites. Next, we use the estimated parameters $\hat{\xi}$ and $\hat{\sigma}$ to  transform the raw exceedances $Y(x)$ observed at site $x$ to exceedances $\tilde{Y}(x)$ with cdf (\ref{eq:censored-distribution}) such that $\xi=1$ and $\sigma=\kappa+1$, i.e., 
 $$\tilde{Y}(x)= (\kappa+1) \left \{ \left( 1+\frac{\hat\xi\, Y(x)}{\hat\sigma}\right)^{1/ \hat\xi}-1 \right \}.$$
Since  $\kappa$ must satisfy
$\Pr(\tilde{Y}(x)>0)=(\kappa+1)^{-1}=0.01$, see Equation~\eqref{eq:marginal}, we get $\kappa=99$.

We fit our hierarchical models to the censored pretransformed data $\tilde{Y}(x)$ by numerically maximizing the pairwise likelihood.  We set the spatial cut-off distance to $\Delta_S=110\,km$, which retains about $60\%$ of the pairs of meteorological stations,  and we choose the temporal cut-off  as  $\Delta_T=10$ hours. The resulting  number of  pairs of observations  is approximately $4.6\times 10^9$, taking into account missing values. The full pairwise likelihood counts around $1.7\times 10^{11}$ pairs, which shows that we have attained a  huge reduction. Pairwise likelihood maximization is coded in C, and it runs in parallel using the R library \texttt{parallel}. All calculations were carried out on a $2.6$ GHz machine with $32$ cores and  $52Gb$ of memory. One evaluation of the composite likelihood requires approximately $18$ seconds. For calculating  standard errors and CLIC$^*$ values,  we use the previously described subsampling technique based on temporal
 windows by considering $B = 500$  overlapping blocks,  each corresponding to $1000$ consecutive hours, i.e. $d_b=50\times 1000$.

We consider  two settings for the hierarchical model,  with (G1) and without  velocity (G2). Then, we compare these two models to 
three variants of a censored Gaussian space-time copula model labeled C1, C2 and C3
\citep{Bortot:Coles:Tawn:2000,Renard:Lang:2007,Davison:Huser:Thibaud:2013}
pertaining to the  
 class  of asymptotic independent processes.
The fits of  the censored Gaussian  space-time copula models  match
a censored Gaussian random field  with transformed threshold exceedances; i.e., we transform original data   to standard Gaussian margins
$G(x)=\Phi^\leftarrow(\text{GP}(\tilde{Y}(x)))$ (with $\Phi$  the standard Gaussian cdf), and we suppose that
$\{G(x),\, x \in \mathcal{X}\}$ is a   Gaussian space-time random field with 
 space-time correlation function $\rho(x_1 , x_2;\theta)$.
 
 We denote by $\rho_e(a)=\exp(-a)$
 and by $\rho_s(a) =  (1 - 1.5 a + 0.5 a^3) 1_{[0,1]}(a)$, $a\geq 0$, the exponential and spherical correlation models with scale $1$, respectively.
We introduce the scaled Mahalanobis distance between spatial locations $s_1$ and $s_2$, written 
 $$a(s_1,s_2;\tau)=\{(s_1-s_2)'\Omega(\tau)^{-1}(s_1-s_2)\}^{1/2}$$
 where
 $$\Omega(\tau)
 =
 \left( 
 \begin{array}{cc}
 \cos(\tau_1)& -\sin(\tau_1)\\
 \sin(\tau_1) &  \cos(\tau_1)\\
 \end{array}
 \right)
 \left( 
 \begin{array}{cc}
 1& 0\\
 0 &  \tau_2^{-1}\\
 \end{array}
 \right)
 \left( 
 \begin{array}{cc}
 \cos(\tau_1)& \sin(\tau_1)\\
 -  \sin(\tau_1) &  \cos(\tau_1)\\
 \end{array}
 \right).
 $$
 The Mahalanobis distance defines elliptical isocontours. Here, $\tau_1\in[0,\pi)$ is the angle with respect to the West-East direction, and $\tau_2>0$ is  the length ratio of the two principal axes.
We choose three specifications of the  space-time  correlation function:
\begin{itemize}
	\item[C1] Space-time separable model:
		\begin{equation}
	\rho(x_1 , x_2;\theta)=\rho_e(a(s_1,s_2;\tau)/\psi_S)\,\rho_e(|t_1-t_2|/\psi_T)\label{eq:sep}
	\end{equation}
with $\theta=(\tau_1,\tau_2,\psi_S,\psi_T)$. We assume anisotropic spatial correlation in analogy to  models G1 and G2. The model is isotropic for $\tau_2=1$.
\item[C2] Frozen field model 1 \citep[see][for a comprehensive account]{Christakos:2017}:
	\begin{equation}
\rho(x_1 , x_2;\theta)=\rho_e(a(s_1-\nu t_1,s_2-\nu t_2;\tau)/\psi)\label{eq:frozen-exp}
\end{equation}
where $\theta=(\tau_1,\tau_2,\psi,\nu')$ and $\nu \in \mathbb{R}^2$ is a velocity vector.
\item[C3] Frozen field model 2 with compact support:
	\begin{equation}
\rho(x_1 , x_2;\theta)=
\rho_s(a(s_1-\nu t_1,s_2-\nu t_2;\tau)/\psi).
\label{eq:frozen-sph}
\end{equation}
In this model, two observations separated
by Mahalonobis distance $a(s_1-\nu t_1,s_2-\nu t_2;\tau)$ greater than $\psi$
will be independent.
\end{itemize}

Evaluation of the full likelihood of the models C1, C2 and C3 requires numerical operations  such as matrix inversion, matrix determinants and high-dimensional Gaussian cdfs \citep{Genz:Bretz:2009},  which are computationally intractable in our case. Therefore, we opt again for a pairwise likelihood approach, which also simplifies model selection through the CLIC$^*$.

Estimation results   are summarized in Table \ref{tab:results}. The CLIC$^*$ in the last column  shows a preference for our hierarchical models with the best value for model G1, followed closely by G2. 
Estimated durations vary only slightly between  G1 and G2. Estimates of $\phi$ differ more strongly, but one has to take into account that estimates of both semi-axis are very close. Moreover, estimates of $\gamma_1 $ and $\gamma_2 $ are similar for  G1 and  G2, which suggests coherent results for the two models and allows reliable physical interpretation of  estimated parameter values. 
Regarding the  results for model G1, we observe  that the estimated parameters $\gamma_1$  and $\gamma_2$ characterize an ellipse covering a large part of the study region, which 
indicates  relatively strong dependence even between sites that are far separated in space.

The estimate of $\phi$ underlines the low inclination of the ellipse, while $\gamma_2 \approx 2 \, \gamma_1$, which leads to an ellongated shape of the ellipse. It corresponds well to the orientation of the mountain ridges in the considered region.

The estimate of $\delta$, which may be interpreted as the average duration of extreme events, correponds well to empirical measures of the actual durations of extreme events 
in the study region. The orientation of the reliefs  seems to play an important role for the estimated velocity characterized by the  values of  $\omega_1$ and $\omega_2$, with $\omega_1$ being considerably larger than $\omega_2$. For visual illustration, Figure~6 shows a simulation of model  G1 where the velocity effect in precipitation intensities becomes apparent. 
This simulation shows heavy precipitation arriving from the north, predominantly spreading over the eastern slopes of a mountain range in the study region, and then becoming more intense and finally gradually evacuating towards the south.

Among the Gaussian copula models, the preference goes to the separable model C1.

To underpin the good fit of our  models through visual diagnostics, Figure \ref{fig:gof} shows estimated probabilities $\Pr(Z(s,t) > q | Z(s',t') > q)$ along different directions and at different temporal lags $|t-t'|$. These plots suggest that the  behavior of models G1 and G2 is very close; there is no strong preference for one model over the other.
The ranking of the  copula models
 based on the CLIC$^*$ is also confirmed by the visual diagnostics.
For contemporaneous observations with time lag $0$, the models C1, C2 and C3 have comparable performance in capturing spatial dependence. However, for lags of $1$ hour, models C2 and C3 represent the space-time interaction not satisfactorily.

Finally, we gain deeper insight into the joint tail  structure of the fitted models
by calculating  empirical estimates $\hat{p}_i(h)$ of the multivariate conditional probability
 $$ 
 \chi^*_{s_i;h}(q) :=\Pr(Z(s_j,t) > q, s_j\in \partial s_i| Z(s_i,t-h) > q)
 $$
 where $\partial s_i$ is the set of the four nearest neighbors of site
 $s_i$, $i=1,\ldots,50$.
We compare  these values with precise Monte-Carlo estimates $\tilde{p}_i^{(j)}(h)$,  $j=1,\ldots,200$, based on a parametric bootstrap procedure  using $200$ simulations of the  models G1, G2 and C1 with the leading CLIC$^*$ values. We compute site-specific root mean squared errors (RMSE)
$$\mathrm{RMSE}_i(h)= \left\{\frac{
 	\sum_{j=1}^{200}(\tilde{p}^{(j)}_i(h)-\hat{p}_i(h))^2}{200}\right\}^{1/2},$$
as well as the resulting total RMSE,  $\mathrm{RMSE}(h)=\sum_{i=1}^{50}\mathrm{RMSE}_i(h)$, as an overall measure of goodness of fit. 
Table \ref{eq:multi-chi} reports such values for fitted models using contemporaneous observations or lags of $1$ or $2$ hours ($h=0,1,2$) between the reference site and its neighbors.
If we consider the quantile $q_{0.99}$ used as a threshold for fitting models,
our hierarchical models present the best fit in terms of RMSE only for lagged values. However,  models G1 and G2 extrapolate better for larger values of the threshold such as  $q_{0.995}$.

%% file: conclusions.tex
\section{Conclusions}\label{sec:conclusions}
We have proposed a novel space-time model for threshold exceedances of data with asymptotically vanishing dependence strength. In the spirit of the hierarchical modeling paradigm with latent layers to capture complex dependence and time dynamics, it is based on   a  latent Gamma convolution process with nonseparable space-time indicator kernels, and therefore amenable to physical interpretation.  This framework leads to marginal and joint distributions that are available in closed form and are easy to handle in the extreme value context. The assumption of conditional independence as in our model is practical since it avoids the need to calculate cumulative distribution functions in large dimensions, although difficulty remains in evaluating the volume of the intersections of more than two cylinders and in calculating partial derivates for full likelihood formula. We can draw an interesting parallel to the max-stable Reich-Shaby process $Z_{RS}(x)$ \citep{Reich:Shaby:2012}, which is one of the more easily tractable spatial  max-stable models and has a related construction. Indeed, the inverted process $1/Z_{RS}(x)$ can be represented as the embedding of a dependent latent convolution process (based on positive $\alpha$-stable variables) for the rate of an exponential distribution. Conditional independence models cannot accurately capture the smoothness of the data generating process. Nevertheless, the $\alpha$-parameter in our model of the Gamma noise in \eqref{eq:gammanoise} partially controls the smoothness of the latent Gamma field $\Lambda(s)$, with smaller values yielding more rugged surfaces.

In cases where data present asymptotic dependence, our asymptotically independent model may substantially  underestimate the probability of jointly observing very high values over several space-time points. Asymptotic dependence in our construction (4) is equivalent to lower tail dependence in $\Lambda(x)$. There is no natural choice for introducing such dependence behavior, but a promising idea is to use what  we label \emph{Beta scaling}: given a temporal process $B(t)$ independent of $\Lambda(s,t)$ with $\mathrm{Beta}(\tilde{\alpha},\alpha)$ distributed margins, $0<\tilde{\alpha}<\alpha$, we could replace $\Lambda(s,t)$ in our construction by the process $\tilde{\Lambda}(s,t)=B(t) \Lambda(s,t)$ possessing margins following the $\Gamma(\tilde{\alpha},\beta)$ distribution. This construction has asymptotic dependence over space, and it will be asymptotically dependent over time if $B(t)$ has lower tail dependence. Follow-up work will explore  theoretical properties and practical implementation of such extensions. 

 We have developed pairwise likelihood inference for our models, which scales well with high-dimensional datasets. We point out that handling observations over irregular time steps and missing data is straightforward with our model thanks to its definition over continuous time.   While we think that MCMC-based Bayesian estimation of the relatively high number of parameters may be out of reach principally due to the very high dimension of the set of latent Gamma variables in the model's current formulation, we are confident that future efforts to tackle the conditional simulation of such space-time processes based on MCMC simulation with fixed parameters could be successful; i.e., by using frequentist estimation of parameters, space-time prediction requires to iteratively update only the latent Gamma field through MCMC, but not parameters.

The  application of our novel model to a high-dimensional real precipitation dataset from southern France was motivated from clear evidence of asymptotic independence highlighted at an exploratory stage. It provides  practical illustration of the high flexibility of our  model and its capability to accurately predict extreme event probabilities for concomitant threshold exceedances in space and time. Based on meteorological knowledge about the precipitation processes in the study region, we had hoped to estimate a clear velocity effect. As a matter of fact,  the fitted  hierarchical model with velocity  appeared to be only slightly superior to other models in some aspects. This interesting finding may also be interpreted as evidence for the highly fragmented structure arising in  precipitation processes at small spatial and temporal scales.  

 Ongoing work aims to adapt the current  latent process construction to the multivariate setting by considering constructions with Gamma factors common to several components, specifically structures with a hierarchical tree-based construction of the latent Gamma components,  and extensions to asymptotic dependence using the above-mentioned Beta-scaling. Ultimately, such novelties could provide a flexible toolbox for multivariate space-time modeling with scenarios of partial or full asymptotic dependence.

%% file: acknowledgment.tex
\section*{Acknowledgment}
The authors express their gratitude towards two anonymous referees and the associate editor for many useful comments that have helped improving earlier versions of the manuscript. The work of the authors was supported by the French national programme LEFE/INSU and by the LabEx NUMEV. Thomas Opitz acknowledges financial support from Ca' Foscari University, Venice, Italy. The authors thank  Julie Carreau  (IRD  
HydroSciences, Montpellier, France) for 
 helping them in collecting the data  from the Meteo France database.

%% file: appendix.tex
\newpage
\section{Appendix 1 : formulas for the Laplace exponent of a random measure}
\label{sec:Laplace}
The Laplace exponent of the random measure $\Gamma(\cdot)$ is defined as
$$
\mathcal{L}(\phi) := -\log\E\left(\exp\left\{-\int \phi(x)\Gamma(dx)\right\}\right)
= \int_{\mathcal X} \log\left\{1+\frac{\phi(x)}{\beta}\right\}\alpha(dx),
$$
where $\phi$ is any positive measurable function. \\
Consider $\phi=v\mathbf{1}_A(x)$. Then, 
 $$
 \mathcal{L}(\phi) = -\log\E\left(\exp\{-v\Gamma(A)\}\right)
 = \int_A\log\left\{1+\frac{v}{\beta}\right\}\alpha(dx)
 = \alpha(A)\log\left\{1+\frac{v}{\beta}\right\},
$$
\emph{i.e.}, $$\displaystyle
\E\left(\exp\{-v\Gamma(A)\}\right)=\left(\frac{\beta}{v+\beta}\right)^{\alpha(A)}.$$
For bivariate analyses, choosing $\phi(x)=v_1\mathbf{1}_{A_1}(x)+v_2\mathbf{1}_{A_2}(x)$, yields 
 \begin{eqnarray*}
   \mathcal{L}(\phi) &=& -\log
 \E\left(\exp\{-v_1\Gamma(A_1)-v_2\Gamma(A_2)\}\right)\\
&=& -\log
 \E\left(\exp\{-v_1\Gamma(A_1\backslash A_2)-(v_1+v_2)\Gamma(A_1\cap A_2)-v_2\Gamma(A_2\backslash A_1)\}\right)\\
 &=& \int_{A_1\backslash A_2}\log\left\{1+\frac{v_1}{\beta}\right\}\alpha(dx)
 + 
 \int_{A_1\cap A_2}\log\left\{1+\frac{v_1+v_2}{\beta}\right\}\alpha(dx)\\
 &&+
 \int_{A_2\backslash A}\log\left\{1+\frac{v_2}{\beta}\right\}\alpha(dx)
 \\
 &= &\alpha(A_1\backslash A_2)\log\left\{1+\frac{v_1}{\beta}\right\}+
 \alpha(A_1\cap A_2)\log\left\{1+\frac{v_1+v_2}{\beta}\right\}+
 \alpha( A_2\backslash A_1)\log\left\{1+\frac{v_2}{\beta}\right\}
 \end{eqnarray*}
and therefore
$$
\E(\exp\{-v_1\Gamma(A_1)-v_2\Gamma(A_2)\})=\left(1+\frac{v_1}{\beta}\right)^{-\alpha(A_1\backslash A_2)}
\left(1+\frac{v_1+v_2}{\beta}\right)^{-\alpha(A_1\cap A_2)}
\left(1+\frac{v_2}{\beta}\right)^{-\alpha(A_2\backslash A_1)}.
$$

\newpage
\section{Appendix 2 : Formulas for the pairwise censored likelihood}
\label{sec:likelihood}

Let $LP^{(1)}(v)$ and $LP_{x,x'}^{(2)}(v_1,v_2)$, $x\ne x'$ denote
 the univariate and bivariate Laplace transform of
$\Lambda(A_x)$  \emph{i.e.}, 
$$LP^{(1)}(v):= \E\left(e^{-v \Lambda(A_x)}\right)=\left(\frac{\beta}{v+\beta}\right)^{c_0},
$$
and
$$
LP_{x,x'}^{(2)}(v_1,v_2):= \E\left(e^{-v_1 \Lambda(A_x)-v_2\Lambda(A_{x'})}\right)=
\left(\frac{\beta}{v_1+\beta}\right)^{c_1}
\left(\frac{\beta}{v_1+v_2+\beta}\right)^{c_2}
\left(\frac{\beta}{v_2+\beta}\right)^{c_3}
$$
with $c_0=\alpha(A_x)$
$c_1=\alpha(A_{x}\backslash A_{x'})$, $c_2=\alpha(A_x\cap A_{x'})$, $c_3=\alpha( A_{x'}\backslash A_x)$.

We obtain
\begin{eqnarray*}
	\frac{\partial}{\partial v}LP^{(1)}(v)&=&- c_0\beta^{c_0}(v+\beta)^{-c_0-1},
	\\
	\frac{\partial}{\partial v_1}LP_{x,x'}^{(2)}(v_1,v_2) &=& 
	-\beta^{c_1+c_2+c_3}\left\{
	c_1(v_1+\beta)^{-c_1-1}
	(v_1+v_2+\beta)^{-c_2}
	(v_2+\beta)^{-c_3}\right. 
	\\
	&&+\left.
	c_2(v_1+\beta)^{-c_1}
	(v_1+v_2+\beta)^{-c_2-1}
	(v_2+\beta)^{-c_3}
	\right\},
	\\
	\frac{\partial}{\partial v_2}LP_{x,x'}^{(2)}(v_1,v_2) &=& 
	-\beta^{c_1+c_2+c_3}\left\{
	c_3(v_1+\beta)^{-c_1}
	(v_1+v_2+\beta)^{-c_2}
	(v_2+\beta)^{-c_3-1}\right. 
	\\
	&&+\left.
	c_2(v_1+\beta)^{-c_1}
	(v_1+v_2+\beta)^{-c_2-1}
	(v_2+\beta)^{-c_3}
	\right\},
	\\
	\frac{\partial}{\partial v_1\partial v_2}LP_{x,x'}^{(2)}(v_1,v_2) &=& 
	\beta^{c_1+c_2+c_3}
	\left\{
	c_1c_2(v_1+\beta)^{-c_1-1}
	(v_1+v_2+\beta)^{-c_2-1}
	(v_2+\beta)^{-c_3}
	\right.
	\\
	&& + \left.
	c_1c_3(v_1+\beta)^{-c_1-1}
	(v_1+v_2+\beta)^{-c_2}
	(v_2+\beta)^{-c_3-1}
	\right.
	\\
	&& + \left.
	c_2(c_2+1)(v_1+\beta)^{-c_1}
	(v_1+v_2+\beta)^{-c_2-2}
	(v_2+\beta)^{-c_3}
	\right.
	\\
	&& + \left.
	c_2c_3(v_1+\beta)^{-c_1}
	(v_1+v_2+\beta)^{-c_2-1}
	(v_2+\beta)^{-c_3-1}
	\right\}.
\end{eqnarray*}